\documentclass{appolb}
\usepackage{epsfig}
\usepackage{amssymb}

\newcommand{\beqa}{\begin{eqnarray}}
\newcommand{\eeqa}{\end{eqnarray}  } 
\newcommand{\beqan}{\begin{eqnarray*}}
\newcommand{\eeqan}{\end{eqnarray*}}
\newcommand{\beq}{\begin{equation}}
\newcommand{\eeq}{\end{equation}}

\def\a{\alpha}
\def\b{\beta}
\def\d{\delta}
\def\D{\Delta}
\def\f{\phi}
\def\g{\gamma}
\def\G{\Gamma}
\def\h{\eta}
\def\k{\kappa}
\def\la{\lambda}
\def\La{\Lambda}
\def\m{\mu}
\def\p{\pi}
\def\r{\rho}
\def\s{\sigma}
\def\S{\Sigma}
\def\t{\tau}
\def\W{\Omega}
\def\x{\chi}
\def\lan{\langle}
\def\ran{\rangle}
\def\hf{\hfill}
\def\ni{\noindent}
\def\hs{\hskip}
\def\vs{\vskip}
\def\taub{{\overline{\tau}}}
\def\etab{{\overline{\eta}}}
\def\cent{\centerline}
\def\AA{A\hs-3.1mm$^{~^\circ}$}

\def\ifmath#1{\relax\ifmmode #1\else $#1$\fi}%
\def\rA{\ifmath{{\mathrm{A}}}}

\def\rb{\ifmath{{\mathrm{b}}}}
\def\rB{\ifmath{{\mathrm{B}}}}
\def\rd{\ifmath{{\mathrm{d}}}}
\def\re{\ifmath{{\mathrm{e}}}}
\def\rf{\ifmath{{\mathrm{f}}}}
\def\rG{\ifmath{{\mathrm{G}}}}
\def\rI{\ifmath{{\mathrm{I}}}}
\def\rK{\ifmath{{\mathrm{K}}}}
\def\rL{\ifmath{{\mathrm{L}}}}
\def\rrm{\ifmath{{\mathrm{m}}}}
\def\rp{\ifmath{{\mathrm{p}}}}
\def\rq{\ifmath{{\mathrm{q}}}}
\def\rR{\ifmath{{\mathrm{R}}}}
\def\rs{\ifmath{{\mathrm{s}}}}
\def\rS{\ifmath{{\mathrm{S}}}}

\def\rT{\ifmath{{\mathrm{T}}}}

\def\rz{\ifmath{{\mathrm{z}}}}
\def\rZ{\ifmath{{\mathrm{Z}}}}
\def\eff{\ifmath{{\mathrm{eff}}}}
\def\ev{\ifmath{{\mathrm{ev}}}}
\def\rms{\ifmath{{\mathrm{rms}}}}

\def\tot{\ifmath{{\mathrm{tot}}}}
\def\out{\ifmath{{\mathrm{out}}}}

\def\side{\ifmath{{\mathrm{side}}}}
\def\BE{\ifmath{{\mathrm{BE}}}}

\def\Qout{\ifmath{Q_{\mathrm{out}}}}
\def\Rout{\ifmath{R_{\mathrm{out}}}}
\def\Qside{\ifmath{Q_{\mathrm{side}}}}
\def\Rside{\ifmath{R_{\mathrm{side}}}}
\def\Qlong{\ifmath{Q_{\mathrm{L}}}}
\def\Rlong{\ifmath{R_{\mathrm{L}}}}

\def\Qslong{\ensuremath{Q^2_\mathrm{L}}}
\def\Qsside{\ensuremath{Q^2_\mathrm{side}}}
\def\Qsout{\ensuremath{Q^2_\mathrm{out}}}

\def\Rlong{\ensuremath{r_\mathrm{L}}}
\def\Rside{\ensuremath{r_\mathrm{side}}}
\def\Rout{\ensuremath{r_\mathrm{out}}}

\def\Rslong{\ensuremath{r^2_\mathrm{L}}}
\def\Rsside{\ensuremath{r^2_\mathrm{side}}}
\def\Rsout{\ensuremath{r^2_\mathrm{out}}}

\def\svec#1{{\mbox{{\footnotesize\bf #1}}}}
\def\vec#1{{\mbox{\bf #1}}}

\begin{document}
\pagestyle{plain}
\newcount\eLiNe\eLiNe=\inputlineno\advance\eLiNe by -1
\title{BOSE-EINSTEIN CORRELATIONS\\ 
IN Z FRAGMENTATION\\
AND OTHER REACTIONS
\thanks{Lecture given at the Cracow School of Theoretical Physics, Zakopane 2001}}
\author{Wolfram KITTEL
\address{HEFIN, University of Nijmegen/NIKHEF, Toernooiveld 1, 6525 ED Nijmegen, The Netherlands}
}
\maketitle

\vs 5mm
\begin{abstract}
Recent experimental studies of Bose-Einstein Correlations in Z fragmentation
are reviewed in view of the need to understand their apparent suppression
for pions originating from different W's. Particular features discussed are 
source elongation, position-momentum correlation, non-Gaussian shape of the 
correlator, transverse-mass dependence, density dependence and dilution, 
space-time shape of the emission function, neutral-pion and 
genuine higher-order correlations.
\end{abstract}

\vs 7mm
\section{Introduction}
As proposed by Hanbury Brown and Twiss \cite{Hanb54} in 1954, the (angular)
diameter 
of stars and radio sources in the universe was successfully determined by 
measuring the intensity correlations between separated telescopes.
Likewise, in particle physics, one can in principle use 
Bose-Einstein correlations between identical bosons to measure the
space-time structure of the region from which particles originate 
in a high-energy collision \cite{Kopy72}, provided these bosons are produced
incoherently.
 
The first experimental evidence for Bose-Einstein correlations
in particle physics dates back to 1959 when, in $\rp\bar \rp$ annihilation at 
1.05 GeV/$c$, Goldhaber et al. \cite{102} observed an 
enhancement at small relative angles in like-sign pion pairs not present 
for unlike-sign pairs. More recently, Bose-Einstein correlations have 
been exploited in hadron-hadron, hadron-nucleus,  nucleus-nucleus, 
e$^+$e$^-$ and lepton-hadron collisions to obtain surprisingly detailed
information on the space-time development of particle production.

The recent revival of interest comes from various directions:\\
\newpage
\noindent
1. Their application to determine the space-time development of a
particle collision.\\
2. Their influence on the measurement of effective masses, in
particular of the W mass at LEP2 \cite{Balle,LonSjo}.\\
3. Their role in the phenomenon of intermittency \cite{bialas,wolf96}.\\
4. Their possible effect on multiplicity 
distribution and single-particle spectra \cite{pratt,biaza,csorgo}.

In this paper, we shall review recent experimental studies on the first point,
in particular for e$^+$e$^-$ collisions leading to hadronic final states at
the Z energy.
We consider this information crucial for an understanding of the
underlying dynamics, and in particular of the apparent suppression of
Bose-Einstein correlations of pions originating from different W's within
the same event.

\section{The correlation formalism}
We start by defining symmetrized inclusive $q$-particle distributions
\beq
\r_q (p_1,\dots,p_q)= \frac{1}{\s_\tot} \frac{\rd \s_q(p_1,\dots,p_q)}{
\prod\limits^q_1 \rd p_q}\ \ ,
\eeq
where $\s_q(p_1,\dots,p_q)$ is the inclusive cross section for $q$
particles to be at $p_1,\dots,p_q$, irrespective of the presence and location
of any further particles, $p_i$ is the (four-) momentum of particle $i$ and
$\s_\tot$ is the total hadronic cross section of the collision under
study.

For the case of identical particles, integration over an interval
$\W$ in $p$-space  yields
\begin{eqnarray}
 &~&\int_\W \r_1(p) \rd p = \lan n\ran\ , \nonumber \\
 &~&\int_\W \int_\W \r_2(p_1,p_2)\rd p_1\rd p_2 =
\lan n(n-1)\ran \ ,\nonumber \\
 &~&\int_\W \rd p_1 \dots \int_\W \rd p_q \r_q (p_1,\dots,p_q) =
\lan n(n-1)\dots (n-q+1)\ran \ ,
\end{eqnarray}
where $n$ is the multiplicity of identical particles within $\W$ in a 
given event and the angular brackets imply the average over the event ensemble.

Besides the interparticle {\it correlations} we are looking for,
the inclusive $q$-particle number densities $\rho_q(p_1,\dots,p_q)$ in general
contain ``trivial'' contributions from lower-order densities. 
It is, therefore, advantageous to consider a new sequence of functions
$C_q(p_1,\dots,p_q)$ as those statistical quantities which vanish whenever one
of their arguments becomes statistically independent of the others.
Deviations of these functions from zero shall be addressed as {\it genuine}
correlations. 

The quantities with the desired properties are the correlation 
functions - also called (factorial) cumulant functions - or, in integrated 
form, Thiele's semi-invariants.\cite{thiele} A formal proof of this property 
was given by Kubo.\cite{kubo}
The cumulant correlation functions are defined as in the cluster expansion
familiar from statistical mechanics via the sequence:
\cite{kahn:uhlenbeck,huang,Mue71}
\begin{eqnarray}
\rho_1(1)& =& C_1(1),\\
\rho_2(1,2)& =& C_1(1)C_1(2) +C_2(1,2),\\
\rho_3(1,2,3)& =& C_1(1)C_1(2)C_1(3)
+C_1(1)C_2(2,3)
+C_1(2)C_2(1,3)
+\nonumber\\
& &\mbox{}
+C_1(3)C_2(1,2)+C_3(1,2,3);
\end{eqnarray}
and, in general, by
\begin{eqnarray}
\rho_m(1,\ldots,m) &=& \sum_{{\{l_i\}}_m}\sum_{\mbox{perm.}}
\underbrace{\left[C_1()\cdots C_1()\right]}_{l_1\,\mbox{factors}}
\underbrace{\left[C_2(,)\cdots C_2(,)\right]}_{l_2\,\mbox{factors}}
 \cdots\nonumber\\
& & \cdots \underbrace{\left[C_m(,\ldots,)\cdots C_m(,\ldots,)
\right]}_{l_m\,\mbox{factors}}.
\label{a:4}
\end{eqnarray}
Here, $l_i$ is either zero or a positive integer and the sets of integers
$\{l_i\}_m$ satisfy the condition
\begin{equation}
\sum_{i=1}^m i\, l_i=m.\label{a:5}
\end{equation}
The arguments in the $C_i$ functions are to be filled by the $m$ possible
momenta in any order. In the above relations we have abbreviated 
$C_q(p_1,\dots,p_q)$ to $C_q(1,2,\ldots,q)$; the summations indicate that all 
possible permutations must be taken (the number under the summation sign 
indicates the number of terms). The sum over permutations is a sum over all
distinct ways of filling these arguments. For any given factor product there
are precisely~\cite{huang}
\begin{equation}
\frac{m!}{
\left[(1!)^{l_1} (2!)^{l_2}\cdots(m!)^{l_m}\right] {l_1!}{l_2!}\cdots{l_m!}}
\label{a:6}
\end{equation}
terms.
    
The relations (\ref{a:4}) may be inverted with the result:
\begin{eqnarray}
C_2(1,2)&=&\rho_2(1,2) -\rho_1(1)\rho_1(2)\ ,\nonumber\\
C_3(1,2,3)&=&\rho_3(1,2,3)
-\sum_{(3)}\rho_1(1)\rho_2(2,3)+2\rho_1(1)\rho_1(2)\rho_1(3)\ ,\nonumber\\
C_4(1,2,3,4)&=&\rho_4(1,2,3,4)
-\sum_{(4)}\rho_1(1)\rho_3(1,2,3)
-\sum_{(3)}\rho_2(1,2)\rho_2(3,4)\nonumber\\
&&\mbox{} +2\sum_{(6)}\rho_1(1)\rho_1(2)\rho_2(3,4)-6\rho_1(1)
\rho_1(2)\rho_1(3)\rho_1(4),
\label{a:4b}
\end{eqnarray}
etc. 
Expressions for higher orders can be derived from the related formulae given
in~\cite{kendall}.
 
It is often convenient to divide  the functions
$\rho_q$ and $C_q$ by the product of one-particle densities, which leads to
the  definition of the  normalized inclusive densities and correlations:
\begin{eqnarray}
R_q(p_1,\dots,p_q) &=& \rho_q(p_q,\dots,p_q)/\rho_1(p_1)\ldots
\rho_1(p_q),\label{3.8}\\
K_q(p_1,\ldots,p_q)& =& C_q(p_1,\ldots,p_q)/\rho_1(p_1)\ldots
\rho_1(p_q).\label{3.9}
\end{eqnarray}
In terms of these functions, correlations have been studied extensively for 
$q=2$. Results also exist for $q=3$, but usually the statistics (i.e. number 
of events available for analysis) are too small to isolate genuine 
correlations. To be able to do that for $q\geq 3$, one must apply moments 
defined via the integrals in Eq.(2), but in limited phase-space cells 
\cite{bialas}.

\section{Alternative views}
\subsection{Pion interferometry}

In Fig.~1 we illustrate the production of two identical pions with momenta
$\vec p_1$ and $\vec p_2$, arising from two sources A and B with 
coordinates $\vec x_\rA$, $\vec x_\rB$ (see also \cite{zalew}). 
The pion wave functions can be written as
\beq
\psi_{1\rA} =  e^{-i \svec p_1\cdot \svec x_\rA + i\a} \ \ \ , \ \ \ 
\psi_{2\rB} =  e^{-i \svec p_2\cdot \svec x_\rB + i\b} 
\label{13-1}
\eeq
if $\p(\vec p_1)$ is emitted from source A and $\p(\vec p_2)$ from source
B, where $\a$ and $\b$ are arbitrary phases of the sources. 

\vs 2mm
\centerline{\epsfig{file=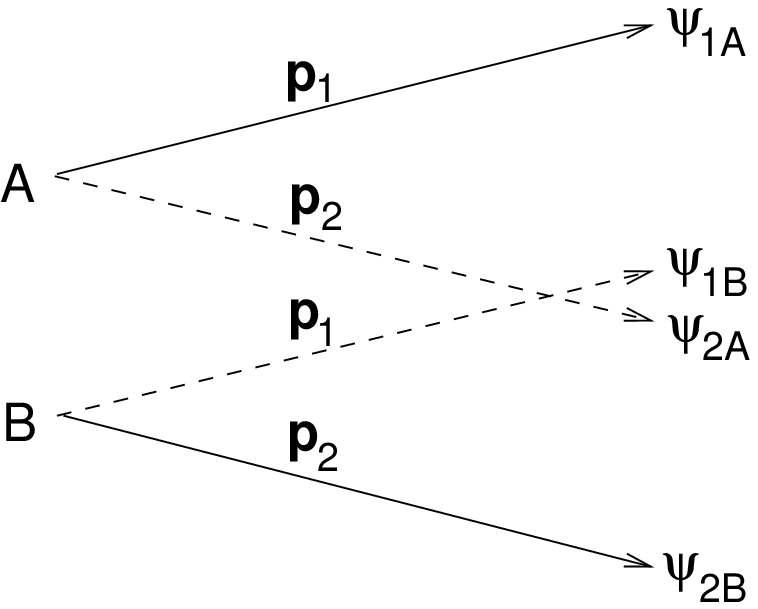,width=5.8cm}}
\vs 2mm
  {\small\baselineskip=12pt\ni
Figure 1. Emission of two identical bosons with momenta $\vec p_1,\vec p_2$
from two sources A, B.\par}

\vs 2mm
If $\p(\vec p_1)$ is emitted from source B and $\p(\vec p_2)$ from source A,
then indices A and B (and the phases) should be interchanged in (\ref{13-1}).
Since the
two pions are identical bosons and the observer cannot
decide from which source a particular pion was emitted, the coincidence 
amplitude for simultaneous observation of two pions with momenta
$\vec p_1$ and $\vec p_2$ has to be Bose-symmetrized:
\vs 2mm
\beq
A_{\BE} = \psi_{1\rA} \psi_{2\rB} + \psi_{1\rB} \psi_{2\rA}\ .
\label{13-2}
\eeq
The corresponding coincidence rate is
\beq
I_{\BE} = |A_\BE|^2=2 + 2 \cos (\D\vec p\cdot\D\vec x)\ ,
\label{13-3}
\eeq
where $\D\vec p=\vec p_1-\vec p_2$ and $\D\vec x=\vec x_\rA-\vec x_\rB$.
Note that the arbitrary phases $\a,\b$ have dropped out from (\ref{13-3}),
which is valid for completely incoherent emission. We define as two-particle
Bose-Einstein ratio $R_2$ the ratio between $I_{\BE}$ and the
rate $I_0$ which would be observed if there were no BE interference:
\beq
R_{2} = I_{\BE}/I_0 = 1 +  \cos (\D\vec p \cdot \D\vec x)\ .
\label{13-4}
\eeq
From (\ref{13-4}) it follows that $R_{2}$ reaches a maximum value of 2
for $\D\vec p=0$.
Furthermore, it can be seen that the momentum difference $\D\vec p$ probes
the source dimensions in a direction parallel to $\D\vec p$. 

We shall, however, see later on that such a simple
picture cannot be maintained due to correlation between $\vec x$ 
and $\vec p$ observed in the data and expected from hydrodynamical models
as well as from string models.
 
One step
more realistic than the binary source considered in Fig.~1 is a source with 
a spherically symmetric Gaussian density distribution of emitting 
centres \cite{102}
\beq
\r(\vec r) \propto \exp \left[ -\vec r^2/(2 r^2_0)\right]\ ,
\label{13-5}
\eeq
which yields as Bose-Einstein ratio
\beq
R_{2} = 1 + \exp \left[ -r^2_0 \D\vec p^2\right]\ .
\label{13-6}
\eeq
or, in its Lorentz-invariant form, 

\beqa
            & R_2(Q^2) & =  1+{\exp}(-r^2_\rG Q^2) \nonumber \\
\mbox{with\hf} &        &  \label{13-7} \\
             & Q^2~   & = -(p_1-p_2)^2=M^2-4m^2_{\p}~~,\nonumber 
\eeqa
where $M$ is the invariant mass of the pion pair.
This corresponds to a Gaussian shape of the source 
in the centre-of-mass system of the pair, where $q_0\equiv\D E=0$.

\subsection{Emission function and Wigner function}

The picture presented in Sub-Sect.~3.1 corresponds to the
one-dimensio\-nal treatment of a spherically symmetric static source.
However, a high-energy collision is neither spherically symmetric nor static.
A formalism particularly handy for the fully-dimensional treatment of 
a dynamical 
emitter is the so-called Wigner-function formalism \cite{Gyu79,pratt90}.
This is based on the emission function $S(x,p)$, a covariant Wigner-transform
of the source density matrix. $S(x,p)$ can be interpreted as a
quantum-mechanical analogue of the classical probability that a boson is 
produced at a given space-time point $x=(t,\vec r)$ with a given 
momentum-energy $p=(E,\vec p)$.

In the general case, the normalized two-particle density $R_2(1,2)$ or 
correlation function $K_2(1,2)$ depend on the momentum components of particles
1 and 2. For the study of correlations, it is conveniant to decompose the two 
single-particle four-vectors $p_1$ and $p_2$ into the 
average $K= [(E_1+E_2)/2,\ \vec K=\vec p_1+\vec p_2)/2]$ 
and the relative momentum $Q=(\D E=E_1-E_2,\vec Q=\vec p_1-\vec p_2)$. 

Starting from the space-time $x$ and momentum-energy $K$ 
dependent pion-emission function $S(x,K)$, the normalized density in 
momentum space can be written as \cite{pratt90}
\beq
R_2(\vec Q,\vec K)\approx1+\frac{|\int d^4x S(x,K)e^{iQ\cdot x}|^2}
{|\int d^4x S(x,K)|^2} = 1+|\lan e^{iQ\cdot x}\ran|^2\ .
\label{12-39}
\eeq
In a Gaussian approximation around the mean space-time production point 
$\bar x$, 
\beq
R_2(\vec Q,\vec K) = 1 + \exp [-Q_\m Q_\nu \lan (x-\bar x)_\m(x-\bar x)_\nu\ran
(\vec K)] + \d R_2 (\vec Q,\vec K)\ . \label{12-40}
\eeq
The variances $\lan(x-\bar x)_\m(x-\bar x)_\nu\ran$ give the size of the 
space-time region from which pions of similar momentum are emitted (which, 
for Gaussian sources, coincides with the more general concept \cite{makh88} 
of {\em lengths of homogeneity}) and $\d R_2$ contains all non-Gaussian 
contributions, usually assumed to be small.

Since the four-momenta $p_i$ of the two particles are on-shell, $Q$ and $K$ 
are in general off-shell but obey the orthogonality and mass-shell constraints
\beq 
Q\cdot K = 0\ ,\ K^2-Q^2/4=m^2\ ,
\eeq
so that only 6 linear combinations of the variances 
are measurable \cite{chap95-2}. If the source is azimuthally symmetric in
coordinate space, a reflection symmetry is present in momentum space with
respect to the plane spanned by $\vec K$ and the event axis. As a
consequence, all mixed variances linear in the direction orthogonal
to this plane (``sidewards'') must vanish and the correlator must be
symmetric under $Q_\side\to-Q_\side$, so that only four linear combinations 
remain measurable!
Note, however, that every one of them depends on $\vec K$.

\subsection{String models}
Alternatively, Bose-Einstein correlations have been introduced
into string models \cite{120,anho86,ander}. In these models, an ordering in 
space-time exists for the hadron momenta within a string.
Bosons close in phase space are nearby in space-time and 
the length scale measured by Bose-Einstein correlations is not
the full length of the string, but the distance in boson-production
points for which the momentum distributions still overlap.

Fig.~2a illustrates the production of (identical) particles 1 and
2 from a color string in $x$ and $t$. The color field breaks up into
quark-antiquark pairs and adjacent quarks and antiquarks recombine
into mesons. The production of the same final state, but with particles
1 and 2 exchanged is described in Fig.~2b.

\begin{figure}[H,t]
\begin{center}
\epsfig{file=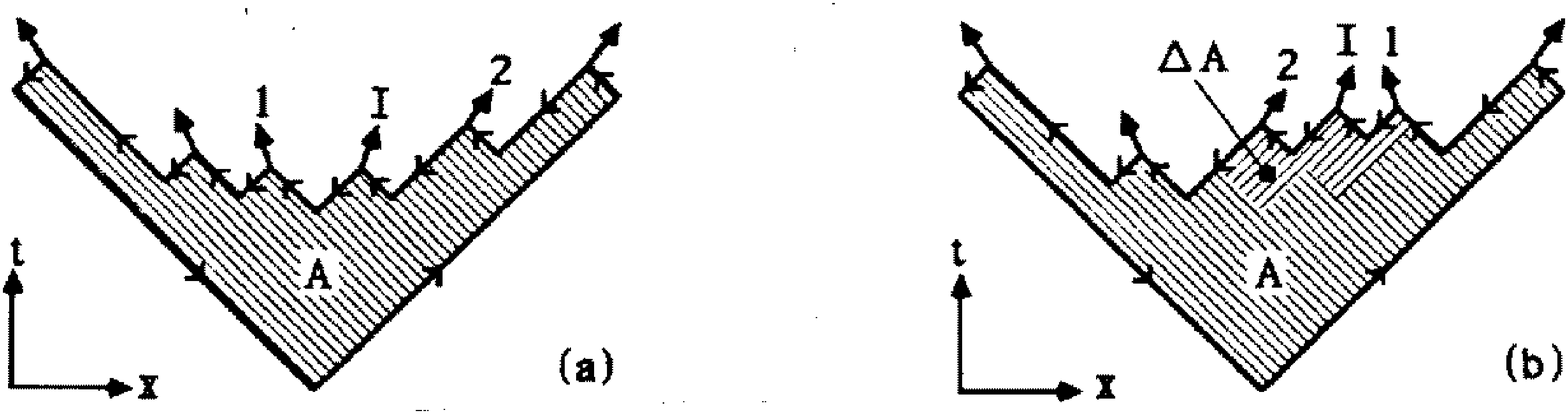,width=12cm}
\end{center}
\vs 2mm
{\small\baselineskip=12pt\ni
Figure 2. Space-time diagram for two ways to produce two identical
bosons in the color-string picture \protect\cite{anho86}.\par}
\end{figure}

In a color-string model, the (non-normalized) probability $\rd\G_n$ to
produce an $n$-particle state $\{ p_j \}$, $j=1,\dots n$ of distinguishable
particles is
\beq
\rd \G_n=[\Pi^n_{j=1} N \rd p_j \d(p^2_j-m^2_j)] \d(\S p_j-P)\exp (-bA_n)\ ,
\eeq
where the exponential factor can be interpreted as the square of a
matrix element 
\beq
M_n = \exp (i\xi A_n)\ , \ \ \ \rR\re (\xi)=\k\ , \ \ \rI\rrm (\xi)=b/2\ ,
\eeq
and the remaining terms describe longitudinal phase space, with $P$
being the total energy-momentum of the state. $N$ is related to the
mean multiplicity and $b$ to the correlation length in rapidity. $A_n$
corresponds to the total space-time area covered by the color field 
(Fig.~2), or to an equivalent area in energy-momentum space divided
by the square of the string tension $\k=1$ GeV/fm \cite{anho86}.

The production of two identical bosons (1,2) is governed by the 
symmetric matrix element
\beq
M=\frac{1}{\sqrt 2} (M_{12}+M_{21})= \frac{1}{\sqrt 2} [\exp (i\xi A_{12})+
\exp (i\xi A_{21})]\ . 
\eeq
From Fig.~2 it is clear that there is an area difference and, 
consequently, a phase difference between $M_{12}$ and $M_{21}$ of
\beq
\D A=|A_{12}-A_{21}|=\frac{1}{\k^2}|\vec p_1 E_2 -\vec p_2 E_1
+ (\vec p_1-\vec p_2) E_\rI - (E_1-E_2)\vec p_\rI |\ , 
\label{13-13b2} 
\eeq
where the indices 1,2 and I represent particles 1, 2 and system I, 
respectively.

Using this matrix element, one obtains
\beq
R_\BE\approx 1+\lan \cos(\k\D A)/\cosh (b\D A/2)\ran\ \ ,
\label{13-13b3} 
\eeq
where the average runs over all I. In the limit $Q^2=-(p_1-p_2)^2=0$,
(\ref{13-13b2}) gives $\D A=0$ and (\ref{13-13b3}) $R_\BE= 2$,
in agreement with the results from the conventional interpretation for
completely incoherent sources. However, for $Q^2\not= 0$ follows an additional
dependence on the momentum $p_\rI$ of the system I produced between
the two bosons.

Corrections to (\ref{13-13b3}) are necessary due to non-zero mass and
transverse momentum of quarks and due to the contribution of 
resonances to the production of particles
of type 1, 2.

The model can account well for most features of the e$^+$e$^-$ data 
\cite{108,109,115}, including the approximately spherical shape of the
BE effect. More recently, the symmetrization has been generalized to more 
than 2 identical particles \cite{anri97}.

\section{Recent experimental results}
\subsection{Existence}
Bose-Einstein correlations are by now a well established effect in the
hadronic final states of Z decay \cite{acto91,abreu94,deca92}.
A clear enhancement is observed in $R_2$ at small $Q$.
This is not a trivial observation since, according to the pion interferometry
interpretation of Sect.~3.1, this would require at least partially chaotic 
pion production.

If present in hadronic Z final states, there is no reason to expect it to
be absent in hadronic W final states (intra-W BEC), and a signal
consistent with that of Z into light quarks is indeed established \cite{L3}.
Examples for $R_2$ as a function of $Q$ are given in Fig.~3,
both for W and Z fragmentation. 

The important question is that of BEC between pions each originating from
a different W in fully hadronic W$^+$W$^-$ final states (inter-W BEC).
If existent, such a correlation would, on the one hand, cause a
potential bias in the mass determination of the W \cite{Balle,LonSjo}. 
On the other,
it could serve as a pion-interferometry laboratory for the measurement
of the space-time development of W fragmentation into pions.
The recent status of the search for inter-W BEC is covered in \cite{Tod}.

\begin{figure}
\begin{center}
\epsfig{figure=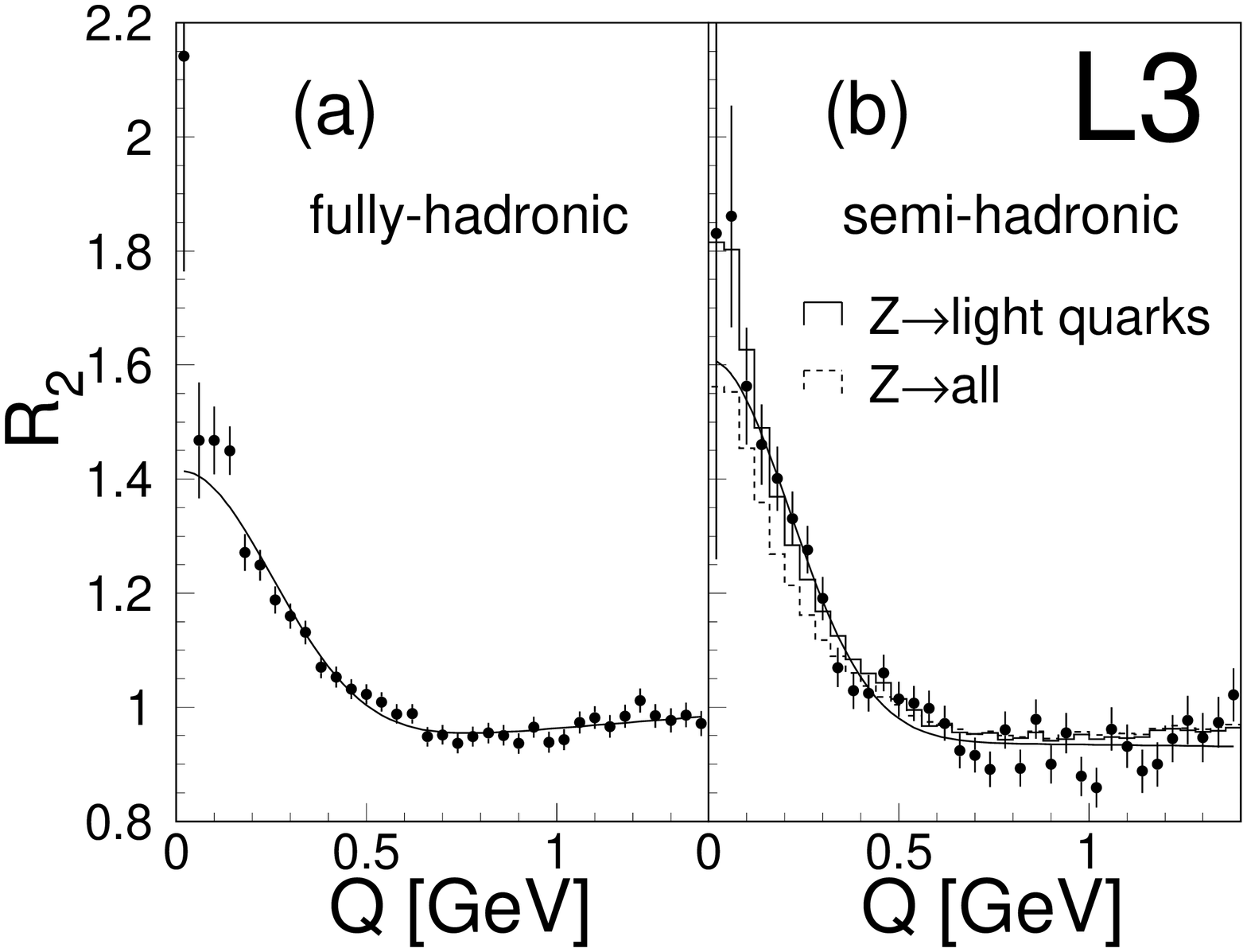,height=7cm}
\end{center}
\vs -4mm
{\baselineskip=12pt\small\ni
Figure 3. The Bose-Einstein correlation function $R_2$ for (a)
the fully-hadronic WW events, and (b) the semi-hadronic WW events. In (b)
the full histrogram is for the light-quark Z decay sample and the dashed
histrogram is for a sample containing all hadronic Z decays. Also shown are 
Gaussian fits to the WW data~\cite{L3}.\par}
\end{figure}

For a detailed understanding of W$^+$W$^-$ overlap and inter-W BEC,
a detailed analysis of BEC in a single W would be required. Given
that only a few thousand of these W's have been produced at
LEP, such a detailed study is presently not possible. It is, however,
possible on the millions of events accumulated at the Z, and we
shall assume that the fragmentation properties are similar for those
two bosons (except for the fact that $\rZ\to \rb\bar\rb$ has no equivalent
in W decay). Since this detailed analysis is still going on, we
shall also look at corresponding properties in hadron-hadron
and even heavy-ion collisions.

\subsection{Pion-source elongation}

The form of the correlation function in more than one dimension
has been a major subject of theoretical study in recent 
years\cite{anri97,ref8,ref4,pratt90,ref6,ref7,csor97}. In Monte Carlo generators, 
spherical symmetry is usually assumed\cite{LonSjo,jadach,wit,kart}, 
while elongation can be expected when a string-like shape is 
maintained\cite{anri97,csor97}.    
Experimentally, detailed three-dimensional analyses were done
for heavy-ion collisions~\cite{heavy1,heavy2} and for hadron-hadron
collisions\cite{na22}. 
While the volume of the pion emission region appeared to be approximately
spherical for heavy-ion collisions, a clear elongation was observed in
hadron-hadron collisions. 
An elongation is now also observed at LEP \cite{L3E,Delphi,Opal}.

In this analysis the longitudinal center-of-mass system 
(LCMS)~\cite{ref6} is used. 
This is defined for each pair of particles as the system,
resulting from a boost along the thrust axis, in which the sum 
of the momenta of the pair is perpendicular to the thrust axis. 
In this system, one can resolve the three-momentum difference of the pair of
particles into a longitudinal component \Qlong\ parallel to the thrust axis, 
\Qout\ along the sum of the particles' momenta (see Fig.~4)
and \Qside\ perpendicular to both \Qlong\ and \Qout. 
Then, the invariant four-momentum difference can be written as~\cite{ref6}
\begin{figure}
\begin{center}
\epsfig{figure=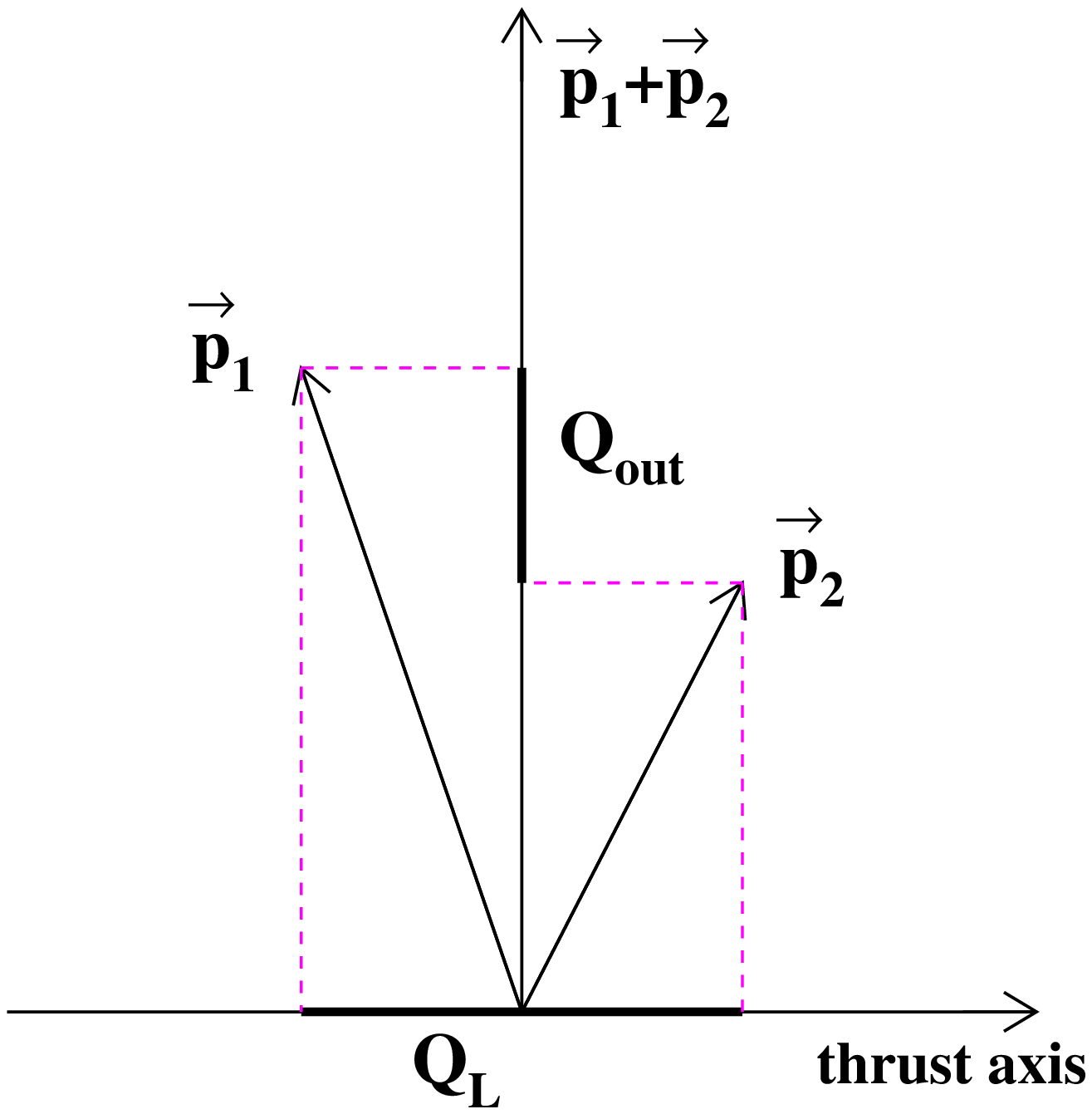,height=6cm}
\end{center}
\vs -2mm
{\baselineskip=12pt\small\ni
Figure 4. The longitudinal center of mass frame (LCMS) showing the projection 
of $Q$ on the (\Qlong-\Qout) plane. \Qside\ is the projection of $Q$ on the 
axis perpendicular to this plane.\par}
\label{lcmsys}
\end{figure}

\begin{equation}
  Q^{2} = \Qlong^2 + \Qside^2 + \Qout^2 -(\Delta E)^2
        = \Qlong^2 + \Qside^2 + \Qout^2 (1-\beta^2),
\end{equation}
where
\begin{equation}
  \beta\equiv\frac{p_{\mathrm{out}\,1}+p_{\mathrm{out}\,2}}{E_{1}+E_{2}}
\label{beta}
\end{equation}
with $p_{\mathrm{out}\,i}$ and $E_{i}\,\,(i=1,2)$ the out-component of 
the momentum and the energy, respectively, of particle $i$ in the LCMS. 
The energy difference $\Delta E$ and therefore the difference in emission 
time of the two particles couples only to the component \Qout.
Consequently, \Qlong\ and \Qside\ reflect only spatial dimensions of the 
source, whereas \Qout\ reflects a mixture of spatial and temporal dimensions.
The correlation function is then parametrized in terms of 
$\vec{Q}=(\Qlong,\Qside,\Qout)$:
\begin{equation}
  R_{2}(\vec{Q})=\frac{\rho_2(\vec{Q})}{\rho_0(\vec{Q})}
    \ \ \ .
\label{eq2}
\end{equation}

Assuming a Gaussian (azimuthally, but not necessarily spherically, symmetric)
shape of the source,
the following three-dimensional parametrization has been 
proposed~\cite{ref4,pratt90,crosst1}:
\begin{eqnarray}
\lefteqn{R_2(\Qlong,\Qout,\Qside) =\qquad\qquad}  \nonumber \\
       & &  \hs-2mm = \gamma \left(1+\delta \Qlong+\varepsilon
          \Qout +\xi \Qside\right)\cdot    \label{eqparcross3}\\
       & & \hs-2mm \cdot \left[1\hs-1mm+\hs-1mm\lambda \exp \left(\hs-1mm-\Rlong^2\Qlong^{2}
         \hs-1mm-\hs-1mm\Rout^2\Qout^2\hs-1mm-\hs-1mm\Rside^2 \Qside^2
         \hs-1mm+\hs-1mm2\rho_\mathrm{L,out}\Rlong\Rout\Qlong\Qout\right)
         \right],\nonumber
\end{eqnarray}
where the factor $(1+\delta \Qlong+\varepsilon \Qout+\xi\Qside)$
takes into account possible long-range momentum correlations in the form
of a slow rise, $\gamma$ is a normalization factor close to unity and
the term between square brackets is the two-particle Bose-Einstein correlation
function associated with a Gaussian shape of the source.

By fitting the correlation function with this parametrization, one can extract
the factor $\lambda$, which measures the strength of the
correlation, and the `radii' $r_i$ ($i =$ L, out and side)
defined as $\sigma_i/\sqrt{2}$, with the $\sigma^{2}_i$ the variances of
a multi-dimensional Gaussian distribution of the source in configuration space.
$\rho_\mathrm{L,out}$ is the correlation between the longitudinal and 
out components of this Gaussian.
In the LCMS, the duration of particle emission only couples to the 
out-direction and only enters in the parameters \Rout\ 
and $\rho_\mathrm{L,out}$.     
Hence, \Rside\ can be interpreted as the transverse component of the 
geometric radius.
The parametrization, Eq.~(\ref{eqparcross3}), assumes azimuthal symmetry 
of the source, which means that the two-particle Bose-Einstein correlation 
function associated with the Gaussian shape of the source, is invariant under 
the transformation $\Qside\rightarrow -\Qside$. Consequently, the only 
possible off-diagonal term is the $\Qlong\Qout$ term. It turns out to be 
zero within errors, however. Note that $\r_{\rL,\out}=0$ is indeed expected
in LCMS near midrapidity or for boost-invariant sources.

The results of three LEP experiments \cite{L3E,Delphi,Opal} are summarized in 
Table 1. In spite of the different selection criteria and reference samples,
all experiments consistently demonstrate an elongated shape of the pion source
(or rather region of homogeneity) in hadronic Z decay. On the other hand, 
$r_\side/r_\rL=1.08\pm0.03$ is found \cite{L3E} for JETSET with BE \cite{jet}.

\renewcommand{\arraystretch}{1.3}
\vs 2mm
\begin{table}[ht]
\vspace*{-0.5cm}
\caption{\small\baselineskip=12pt
Elongation of the pion source in hadronic $Z^0$ decays:
  summary of the measurements at LEP1 (\protect$r_{\rT}$ corresponds to
  \protect $Q_{\rT}=\sqrt{Q_{\out}^2+Q_{\side}^2}$).\label{tab:2dim}}
\begin{center}
\footnotesize
\begin{tabular}{|l|c|c|c|}
\hline
\begin{minipage}{1cm}
\begin{center}
$\ $\\
$\ $\\
$\ $\\
$\ $
\end{center}
\end{minipage}
&\begin{minipage}{2.8cm}
\begin{center}
L3\\
``mixed'' reference, \\all events
\end{center}
\end{minipage}
&\begin{minipage}{2.8cm}
\begin{center}
DELPHI\\
``mixed'' reference, \\ 2-jet events
\end{center}
\end{minipage}
&\begin{minipage}{2.3cm}
\begin{center}
OPAL\\
``$+-$'' reference, 2-jet events
\end{center}
\end{minipage}
\\ \hline
$\lambda$&$0.41\pm 0.01^{+0.020}_{-0.019}$&$0.261\pm 0.007\pm 0.010$&$0.443\pm 0.005$\\ \hline
$r_{\rL}$, fm&$0.74\pm 0.02^{+0.04}_{-0.03}$&$0.85\pm 0.02\pm 0.07$&$0.989\pm 0.011^{+0.030}_{-0.015}$\\ \hline
$r_{\out}$, fm&$0.53\pm 0.02^{+0.05}_{-0.06}$&&$0.647\pm 0.011^{+0.024}_{-0.124}$\\ \hline
$r_{\side}$, fm&$0.59\pm 0.01^{+0.03}_{-0.13}$&&$0.809\pm 0.009^{+0.019}_{-0.032}$\\ \hline
$r_\side/r_\rL$&$0.80\pm 0.02^{+0.03}_{-0.18}$& &$0.818\pm 0.018^{+0.008}_{-0.050}$\\ \hline
$r_{\rT}$, fm&&$0.53\pm 0.02\pm 0.07$&\\ \hline
$r_{\rT}/r_{\rL}$& &$0.62\pm 0.02\pm 0.05$&\\ \hline
\end{tabular}
\end{center}
\vspace*{-0.5cm}
\end{table}

A systematic study of the hierarchy of radii obtained from JETSET was
recently performed in \cite{Fia}. Starting from a spherically symmetric
Gaussian correlator, the authors obtain 
$$r_\side>r_\rL> r_\out\ \ ,$$
unlike the experimentally observed elongation, both when using momentum
shifting \cite{jet} or event weighting \cite{Bia} to simulate
the correlation. Generalizing to asymmetric weights, the experimentally
observed elongation can be reproduced, but finding a good set of input
radius parameters turns out an involved procedure.

What is important to realize is that the measured longitudinal radius has
nothing to do with the elongation of the $\rq\bar\rq$ string stretched in 
Z decay. We shall see in Sect.~4.6 that the full pion-emission function is
of the order of 100 fm long, so that $r_\rL$ only measures a small fraction
of it. The reason for that is a strong $\vec x, \vec p$ correlation. Pions
produced at a large distance on the string also have very different
momenta and do not correlate. The ``radii'' therefore measure the effective
size of the source segment radiating mesons with sufficiently small relative
momentum (length of homogeneity), as shown in Fig.~5.

\begin{figure}[h,t,b]
\cent{\epsfig{file=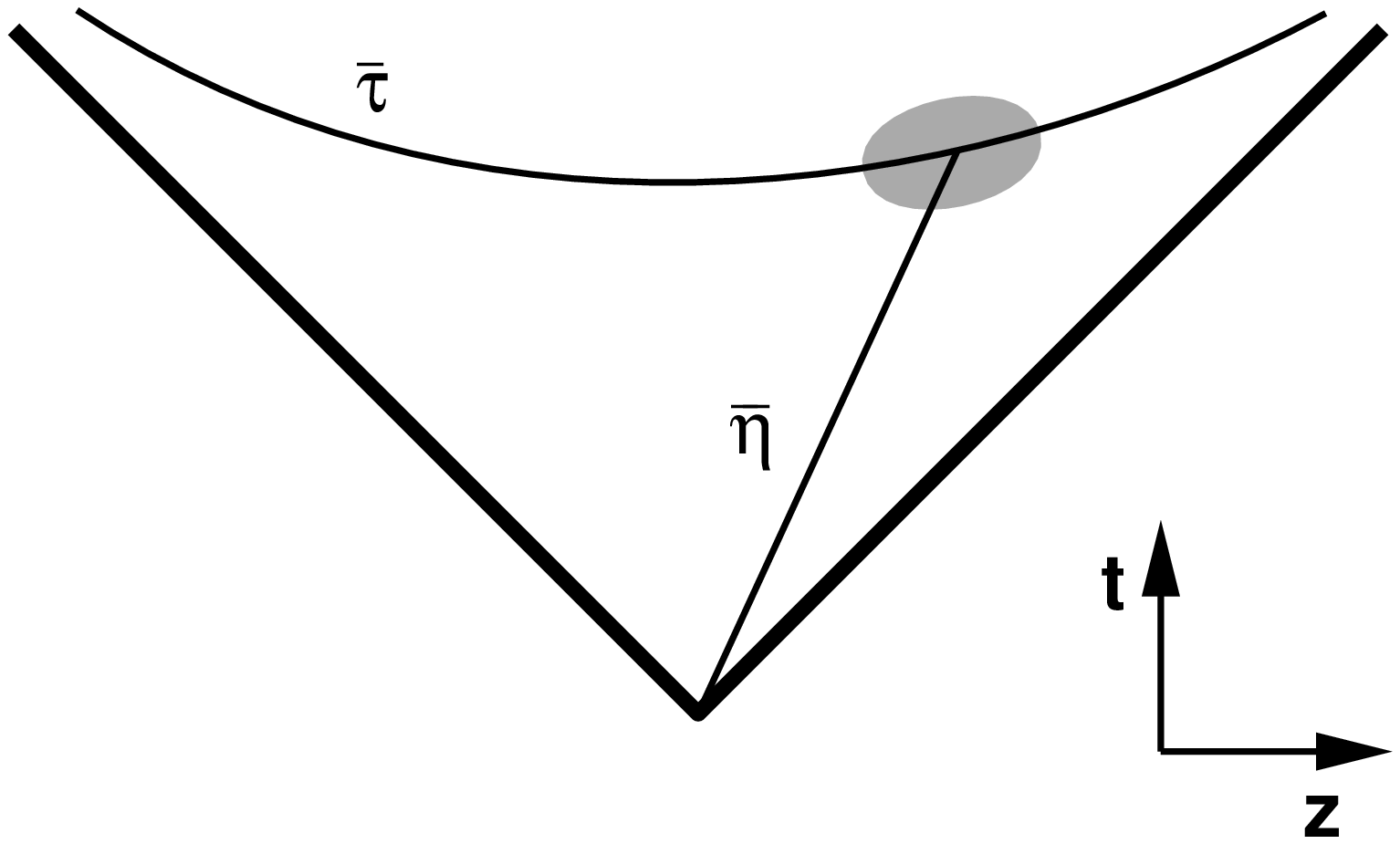,width=7cm}}
\vs 2mm
{\small\baselineskip=12pt\ni
Figure 5. Space-time picture of particle emission for
a given fixed mean momentum of the pair.
The mean value of the proper-time
and the  space-time rapidity    distributions
is denoted by $\taub$ and  $\etab$.
As the rapidity of the produced particles changes from the smallest
to the largest possible value, 
the $[\taub(y),\etab(y)]$ variables scan the
surface of mean particle production in the $(t,z)$   plane~\cite{csor97}.
\par}
\end{figure}

\subsection{The functional form of the correlation function}

More important than the parameters extracted from ``forcing'' the 
two-particle correlation function into a fit by a pre-selected 
parametrization, is the actual experimentally observed shape of this
distribution, itself. 

The simple geometrical interpretation of the interference pattern
based on the optical analogy as in Sect. 3.1 is invalid when emitters
move relativistically with respect to each other, leading to strong 
correlations between the space-time and momentum-energy coordinates of
emitted particles \cite{koleh86,bowl91}. Correlations of this type
arise due to the nature of inside-outside cascade dynamics \cite{bjor83} as
in colour-string fragmentation \cite{artru74}. In the interpretation
of BEC by Andersson and Hofmann \cite{anho86}
in the string model, the length scale measured by BEC is therefore
not related to the size of the total pion emitting source, but to the
space-time separation between production points for which the momentum
distributions still overlap. This distance is, in turn, related to the
string tension. The model predicts an approximately exponential shape
of the correlation function
\beq
R_2(Q) = R_0(1+\la\exp(-r Q))\ ,
\label{13-68}
\eeq
where $r$ is expected to be independent of the total interaction energy.

Furthermore, scale-invariant dynamics is strongly connected
with Bose-Einstein correlation. Scale invariance implies that multiparticle
correlation functions exhibit power-law behavior over a considerable
range of the relevant relative distance measure (such as $Q^2$) in phase
space \cite{bialas,wolf96}. As such, BEC from a static source do not exhibit 
power-law behavior. However, a power law is obtained if the size of the 
particle source fluctuates event-by-event, and/or, if the source itself is a 
self-similar (fractal-like) object extending over a large volume \cite{bia92}.
In these studies, the ratio $R_2$ is parametrised using the form
\beq
R_2(M)=A+B \left( \frac{1}{Q^2}\right)^\b \ .
\label{13-69}
\eeq

\begin{figure}[t]
\cent{\epsfig{file=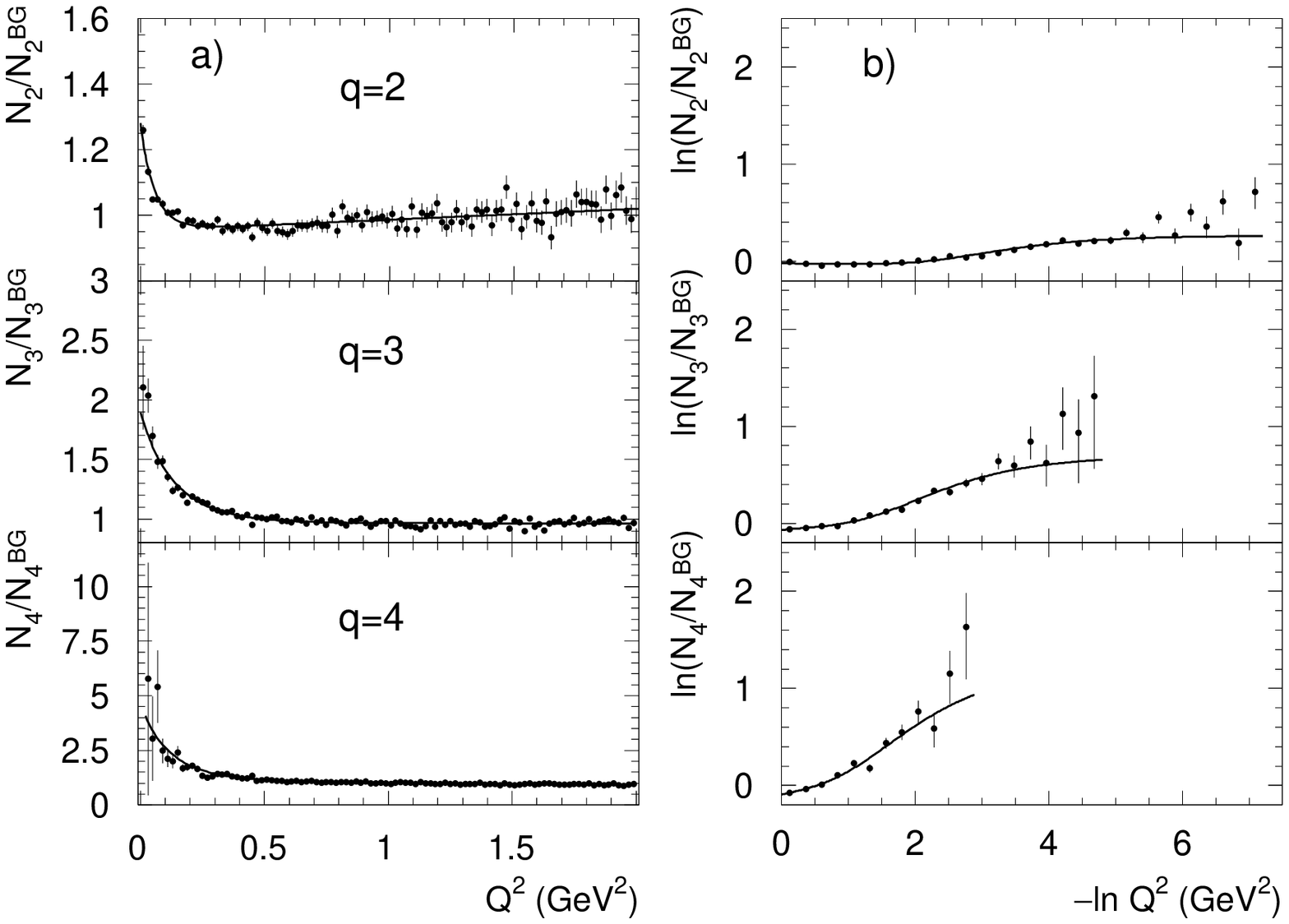,width=12.5cm}}
\vs 2mm
{\small\baselineskip=12pt\ni
Figure~6.  The normalized two-, three- and four-particle inclusive
densities as a function of $Q^2$ (left) and -ln$Q^2$ (right)~\cite{agab95}.
Curves show the multi-Gaussian fits according to \cite{n}.\par}
\end{figure}

The usually ``reasonable'' $\chi^2$ values of the Gaussian fits hide the fact 
the Gaussian parametrization in general fails at low values of $Q^2$, where
statistical errors are often large. For the case of two-particle correlations,
this has been demonstrated convincingly by NA22 \cite{agab93} and UA1 
\cite{neu93}, but deviations from a Gaussian are also observed in 
lepton-hadron \cite{adams93,adloff97} and $\re^+\re^-$ \cite{abreu94} 
collisions. 

This failure of a (multi-) Gaussian form persists in higher-order 
correlations. In Fig.~6a the NA22 data~\cite{agab95} on BE correlations 
of order $q=2$ to 4 are plotted as a function of $Q^2$, in conventional 
linear scale. The curves are the fits by a 
$q$-fold Gaussian parametrization \cite{n}. In Fig.~6b the same data 
and the same fits are repeated for $Q^2<1$ GeV$^2$ on ln-ln scale (where 
the $Q^2$ axis is reflected, i.e., small $Q^2$ correspond to large 
$-\ln Q^2$). Even though the statistical errors at small $Q^2$ are large 
(the very reason why small $Q^2$ 
does not contribute much to $\chi^2$), it is obvious that small-$Q^2$ points 
{\it systematically} lie higher than the multi-Gaussian fit, thus supporting a 
power-law behavior. This effect is even enhanced when the data are 
corrected for Coulomb repulsion. 

\begin{figure}
\begin{minipage}[t]{6cm} a) \end{minipage}\hs 1cm
\begin{minipage}[t]{6cm} b) \end{minipage}
\centerline{\epsfig{file=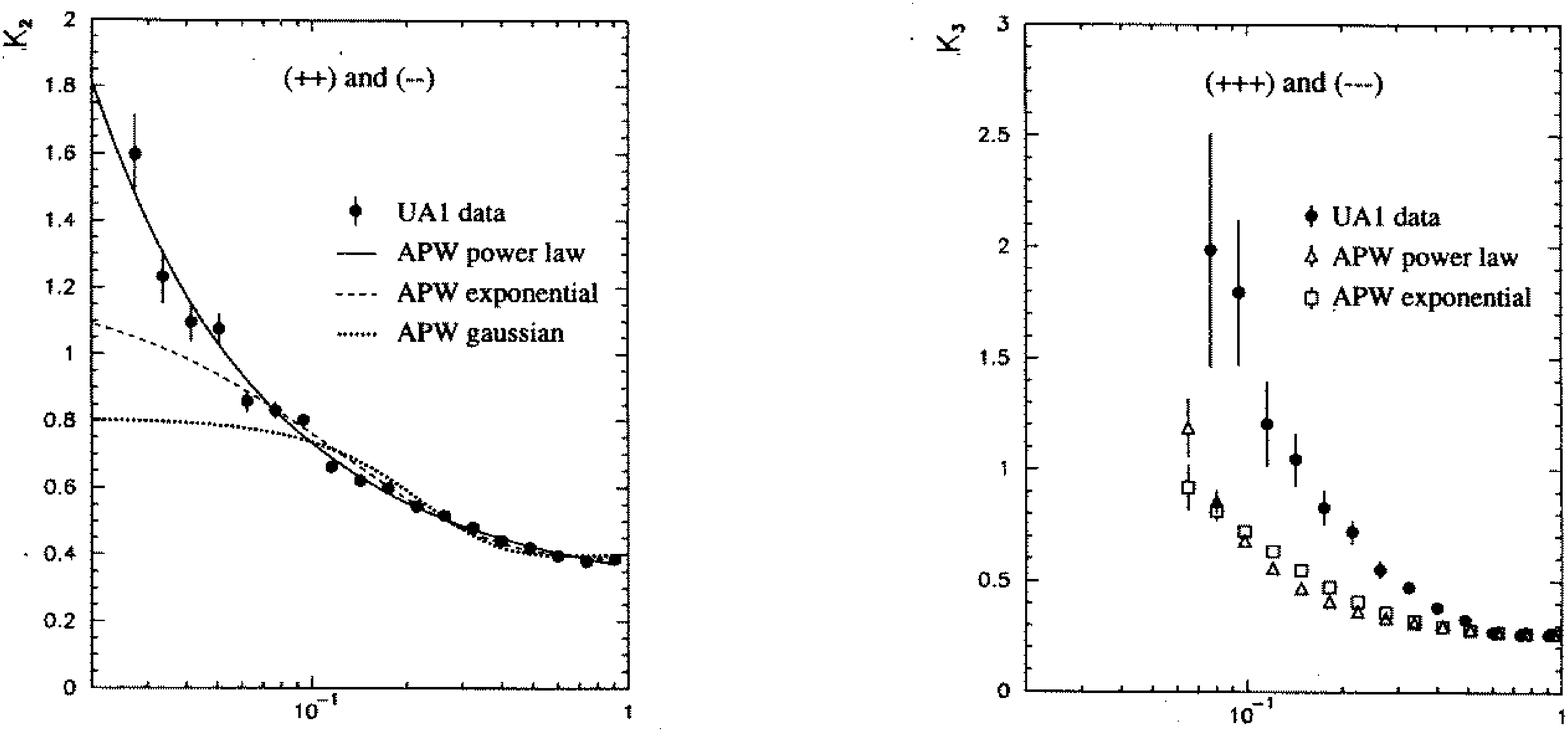,width=12.5cm}}
\begin{minipage}[t]{5cm} {\sf\small\hf Q, GeV} \end{minipage}
\begin{minipage}[t]{7.2cm} {\sf\small\hf max(Q$_{12}$,Q$_{23}$,Q$_{31}$), GeV}
 \end{minipage}
\vs 2mm
{\small\baselineskip=12pt\ni
Figure~7. a) Second-order cumulant with fits of the forms given.
b) Third-order cumulant with APW  predictions based on $K_2$ \cite{egge97}.
\par}
\end{figure}

Fig.~7a shows~\cite{egge97} 
the second-order cumulant $K_2$ as a function of $Q$ 
(on log-scale) compared to a general quantum statistical model,
based on a classical source current formalism applied successfully in
quantum optics \cite{apw}. It includes as special cases more specific
models such as \cite{Gyu79} and \cite{n}. The APW normalized cumulant
predictions are built from normalized correlators $d_{ij}$, the on-shell
Fourier transforms of classical space-time current correlators.
The specific parametrizations tested in Fig.~7 are
\beqa
 & \mbox{Gaussian :} &   d_{ij} = \exp(-r^2 Q_{ij}^2) \nonumber \\
 & \mbox{exponential :} &   d_{ij} = \exp(-r Q_{ij})  \\
 & \mbox{power law :} &   d_{ij} = Q_{ij}^{-\a} \ .\nonumber 
\label{13-70}
\eeqa
For constant chaoticity $\la$ and real-valued currents, APW predict as
second- and third-order normalized cumulants
\beq
K_2(Q_{12})=2\la(1-\la)d_{12}+\la^2d^2_{12} \ , 
\eeq
\beq
K_3 (Q_{12}, Q_{23}, Q_{31}) = 2\la^2(1-\la) 
[ d_{12}d_{23}+d_{23}d_{31}+d_{31}d_{12} ] 
+2\la^3 d_{12}d_{23}d_{31}\ . 
\eeq
The fits in Fig.~7a contain an additive ``background'' parameter, in
addition to $\la$ and $r$ as free parameters. The Gaussian fit is clearly 
excluded and the best fit is achieved with the power-law parametrization
of the correlation function (full line).

Also $K_3$ plotted in Fig.~7b shows a power-law increase. It is, 
furthermore, visible in Fig.~7b that the increase is faster than 
expected from APW $K_2$, even for the power-law parametrization. 

So, there is ample room for improvement of the models and we believe that 
the recently developed methods of studying the correlations (higher-order 
cumulants, higher dimensionality, alternative parametrizations of the 
correlation function) have opened the 
way for an improvement of these models.

An interesting extension of the usual Gaussian approximation
of the BE correlation function is an Edgeworth expansion \cite{kendall}
as suggested in \cite{heg91}
\beq
R_2(Q) = \g(1+\la^*\exp(-t^2/2) [1+\frac{\kappa_3}{3!} H_3 (t) + 
\frac{\kappa_4}{4!} H_4 (t) + \dots ])\ ,
\eeq
with $t=\sqrt 2 Q\cdot r$, $H_n$ being the $n$-th Hermite polynomial,
and $\kappa_n$ the $n$-th order cumulant moment of the correlation
function, where $\kappa_2$ yields $r$. The Hermite polynomials of odd
order vanish at the origin, so that 
\beq
\la = \la^* [1 + \kappa_{4}/8 + \dots ]\ . 
\eeq

A generalization to higher dimensions is straightforward \cite{heg91},
except for possible correlations between the $Q_i$ variables.

The influence of the non-Gaussian shapes was studied \cite{heg91}
on AFS \cite{x,kul90}, E802 \cite{abbo92} and NA44 \cite{lor93} data.
In Fig.~8, a $Q_\rT$ projection of a 2D Edgeworth fit is compared to
that of a 2D Gaussian fit to the E802 data. The deviation from a Gaussian
(dashed) is obvious, and the Edgeworth expansion (full line) is
flexible enough to describe it (with $\la=1!$).

\begin{figure}
\cent{\epsfig{file=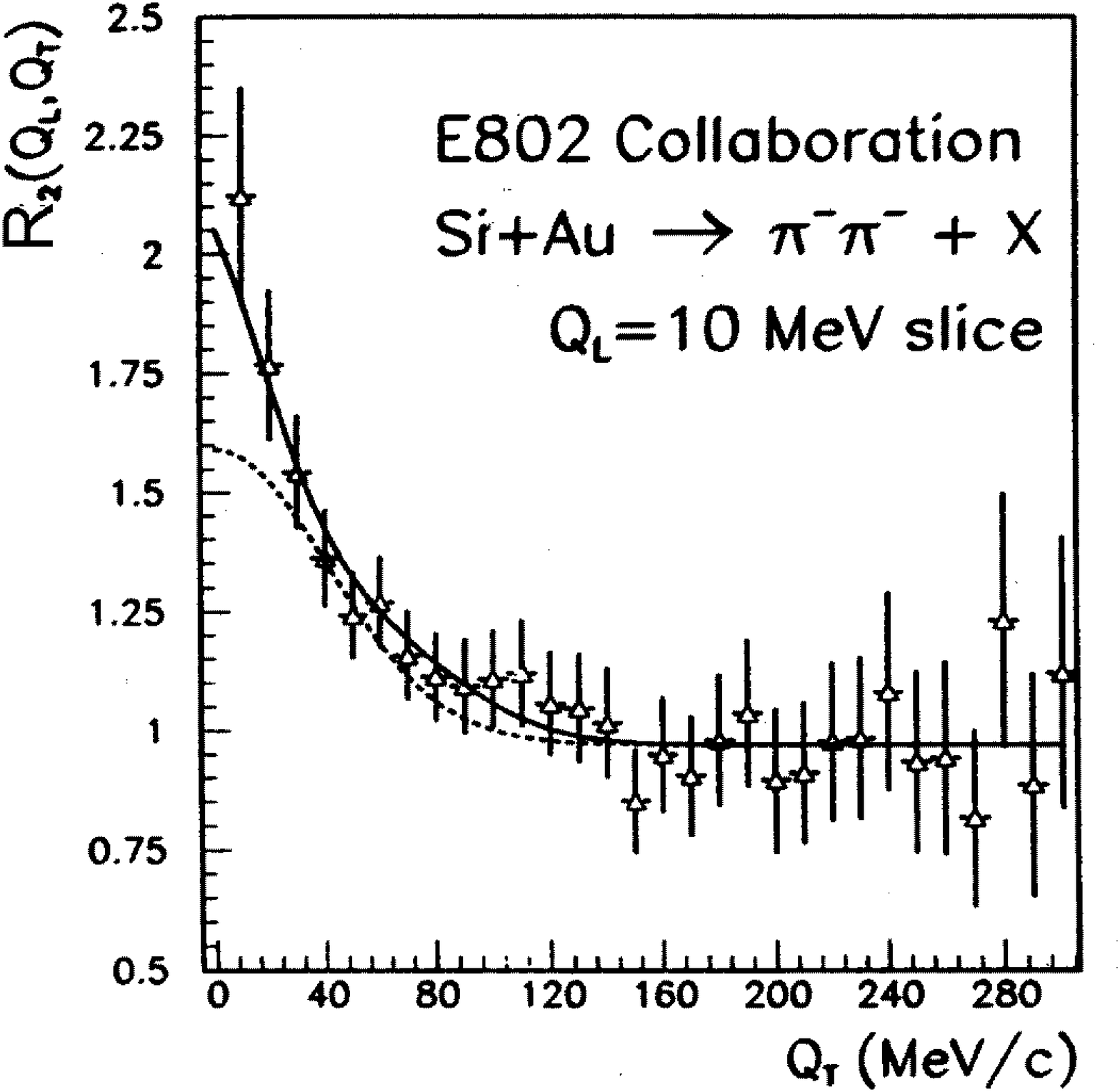,width=6.5cm}}
{\small\baselineskip=12pt\ni
Figure 8. Projection onto $Q_\rT$ of a two-dimensional Gaussian (dashed) 
and Edgeworth (solid) fits for a small-$Q_\rL$ slice \cite{heg91}.
\par}
\end{figure}
 
In Fig.~7a it was shown that even an exponential is not steep enough to
reproduce the fast increase of $K_2$. An interesting observation of
\cite{heg91} is, that a Laguerre expansion of an exponential can
reproduce these UA1 and the NA22 data (Fig.~9). However, at low $Q^2$
data are still systematically above the fit and a power-law fit is
reported in \cite{heg91} to give similarly good $\x^2$/NDF with a 
smaller number of fit parameters. With a core-halo model \cite{csor96} 
strength parameter of $\la_*=1.14\pm 0.10$ (UA1) and $\la_*=1.11\pm 0.17$ 
(NA22), i.e.,  at maximum possible value (unity), there are either other than BE correlations at
work or all resonances are resolved at these low $Q^2$ values.
This may imply the connection between the observed power-law behavior 
(intermittency) and resonance contributions of BE correlations \cite{bia92}.

\vs 5mm
\cent{\epsfig{file=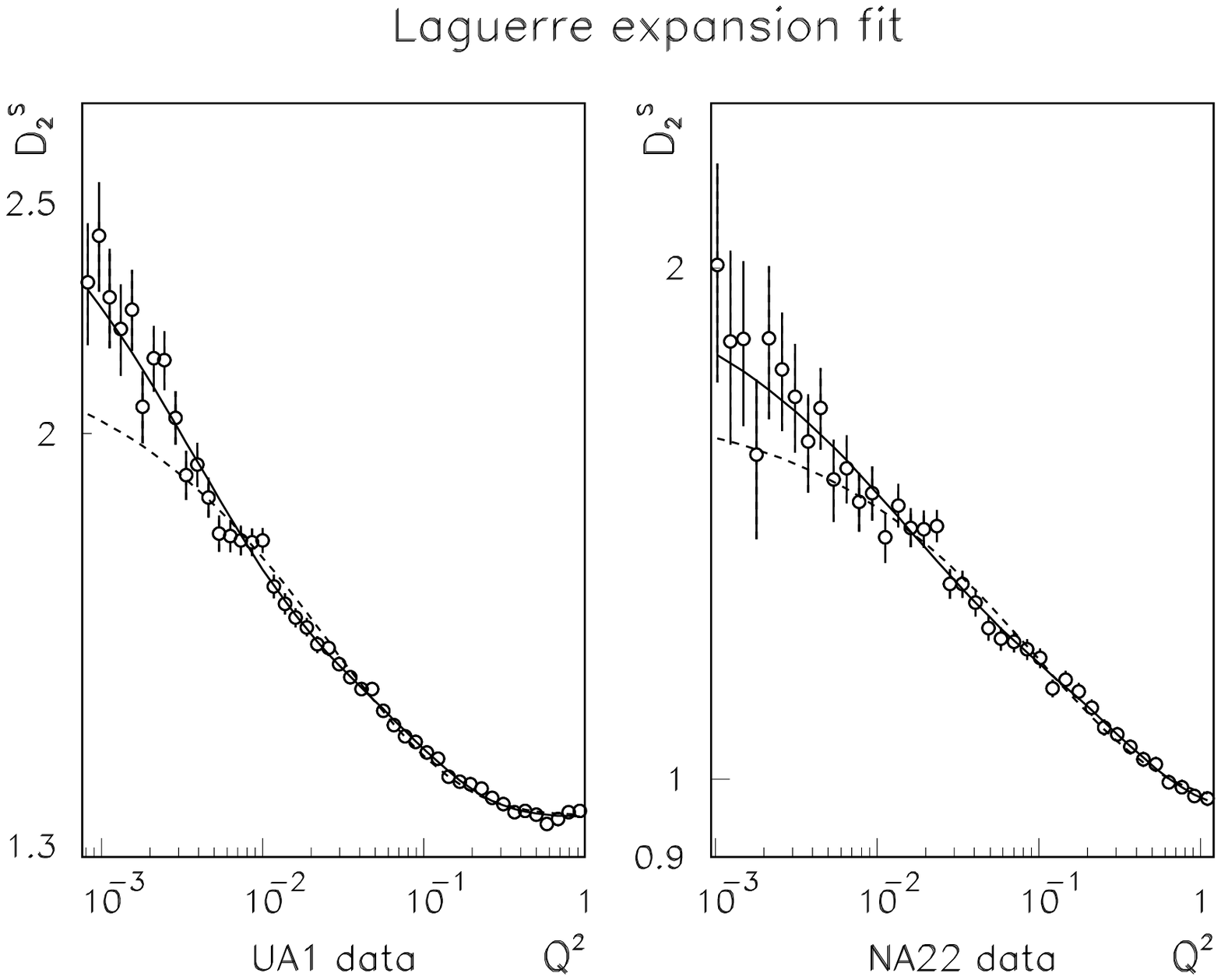,width=12.5cm}}
\vs 2mm
{\small\baselineskip=12pt\ni
Figure~9. The figures show $F^s_2$ which is proportional to the two-particle
Bose-Einstein correlation function, as measured by the UA1~\cite{neu93} 
and the NA22~\cite{agab93}
Collaborations. The dashed lines stand for the exponential fit, 
the solid lines for that with the Laguerre expansion \cite{heg91}. 
\par}
\vs 4mm

In a 3D analysis of e$^+$e$^-$ collisions at the Z-mass \cite{L3E},
results more satisfactory than those obtained with either Gaussian or 
exponential parametrizations were obtained with the Edgeworth expansion.
Taking only the lowest-order non-Gaussian term into account (and dropping
the off-diagonal term), Eq.(30) results in
\begin{eqnarray}
  \lefteqn{R_2(\Qlong,\Qout,\Qside) =\qquad\qquad}  \nonumber \\
       & &   \gamma \left(1+\delta \Qlong+\varepsilon
          \Qout +\xi \Qside\right)        \nonumber \\
       & & \cdot\left\{ \vphantom{\left[ \frac{\kappa_\mathrm{L}}{3!}  \right] }
           1+\lambda\exp \left( -\Rslong\Qslong
               -\Rsout\Qsout-\Rsside\Qsside\right)  \right.  \\
       & & \left.
           \cdot\left[1\hs-1mm+\hs-1mm\frac{\kappa_\mathrm{L}}{3!}H_{3}
(\Rlong\Qlong)\right]
           \left[1\hs-1mm+\hs-1mm\frac{\kappa_\mathrm{out}}{3!}H_{3}
(\Rout\Qout)\right]
           \left[1\hs-1mm+\hs-1mm\frac{\kappa_\mathrm{side}}{3!}H_{3}(\Rside\Qside)\right]
           \right\} \ , \nonumber
  \label{paredge}
\end{eqnarray}
where 
$\kappa_i$ ($i=\mathrm{L,out,side}$) is the third-order cumulant moment
in the corresponding direction and 
$H_{3}(R_{i}Q_{i})\equiv (\sqrt{2}R_{i}Q_{i})^{3}-3\sqrt{2}R_{i}Q_{i}$ 
is the third-order Hermite polynomial.
Note that the second-order cumulant corresponds to the radius $r_{i}$. 
Applying this expansion to the L3 data \cite{L3E} discussed in Sect.~4.2 
improves the confidence level of the fit from 3\% to 30\%.
Non-zero values of the $\kappa$ parameters indicate the deviation from 
a Gaussian, $\lambda$ is larger than the corresponding Gaussian $\lambda$  
and the values of the radii confirm the elongation observed from the Gaussian 
fit. 

\subsection{(Transverse) mass dependence}

\subsection*{4.4.1. The K$^\pm$K$^\pm$ system}

Kaons are less affected by resonance decay than pions and could
eventually provide a cleaner signal of the source. Bose-Einstein
correlations among equally-charged kaons were observed in hh
\cite{107,agui92}, AA \cite{NA44-2,akib93} and e$^+$e$^-$
\cite{abreu96,OPA} collisions (see Table 2 for the latter).

\begin{table}[tb]
\begin{center}
{\footnotesize
\begin{tabular}{|c|c|c|c|c|c|}
\hline
Pair & $\la_\rG$  & $r_\rG$ fm & Ref. & Selection & Ref. \\
     &            &            &      &           & sample\\
\hline
$\p^\pm\p^\pm$     & $0.35\pm0.04$   & $0.42\pm0.04$ & DELPHI \cite{abreu94} 
                   & 2-jet  & mixed \\
                   & $0.40\pm0.02$   & $0.49\pm0.02$   & ALEPH \cite{deca92} 
                   & 2-jet  & mixed \\
                   & $0.58\pm0.01$   & $0.79\pm0.02$   & OPAL \cite{alex96} 
                   & all  & MC \\
\hline
$\p^\pm\p^\pm$   & $0.45\pm0.02$ & $0.82\pm0.03$ & DELPHI \cite{abreu94} 
                 & all  & unlike \\
                 & $0.62\pm0.04$ & $0.81\pm0.04$ & ALEPH \cite{deca92} 
                 & 2-jet & unlike \\
                 & $0.67\pm0.01\pm0.02$ & $0.96\pm0.01\pm0.02$ & OPAL 
   \cite{alex96} & all  & unlike \\
                 & $0.65\pm0.02$ & $0.91\pm0.01$ & OPAL \cite{alex96}
                 & 2-jet & unlike \\
\hline
$\p^\pm\p^\pm$     & $1.06\pm0.05\pm0.16$ & $0.49\pm0.01\pm0.05$ & 
DELPHI\cite{abreu94} & prompt  & \\
                   &      &      &      &  pions & \\  
\hline
K$^\pm$K$^\pm$      & $0.82\pm0.11\pm0.25$ & $0.48\pm0.04\pm0.07$ & 
DELPHI \cite{abreu96} & all & unlike \\
                & $0.82\pm0.22^{+0.17}_{-0.12}$ & $0.56\pm0.08^{+0.08}_{-0.06}$ & OPAL \cite{OPA} & 2-jet & mixed \\
\hline
K$^0_\rS$K$^0_\rS$ & $1.14\pm0.23\pm0.32$ & $0.76\pm0.10\pm0.11$ & 
OPAL \cite{acto93} & all & MC \\
                   & $0.61\pm0.16\pm0.16$ & $0.55\pm0.08\pm0.12$ & 
DELPHI \cite{abreu96} & all & MC \\
                   & $0.96\pm0.21\pm0.40$ & $0.65\pm0.07\pm0.15$ & 
ALEPH \cite{busk94} & all & MC \\
\hline
\end{tabular}
\par}
\end{center}
\vs 2mm
{\small\baselineskip=12pt\ni
Table 2. Parameters $\la_\rG$ and $r_\rG$ in the Gaussian parametrization in
e$^+$e$^-$ interactions at LEP, for different like-charged particles 
\cite{OPA}.\par}
\end{table}

The size of the kaon emission region tends to be smaller than that of the 
pion emission region, in particular in AA collisions. The difference in 
resonance effects on $\p\p$ and KK correlations only partially can explain 
this difference in AA collisions. In e$^+$e$^-$ collisions, the
Gaussian radius parameter $r_\rG$ tends to
be smaller for K$^\pm$K$^\pm$ than for $\p^\pm\p^\pm$, but the spread
is large due to different choices of background.

\subsection*{4.4.2. The K$^0_S$K$^0_S$ system}

The K$^0_\rS$K$^0_\rS$ system is a mixture of K$^0\bar\rK^0$ and
$\rK^0\rK^0$ $(\bar\rK^0\bar\rK^0)$ pairs. 
At LEPI energy, only 28\% of all $\rK^0_\rS\rK^0_\rS$ pairs
are estimated to come from the (identical) $\rK^0\rK^0$ or
$\bar\rK^0\bar\rK^0$ system.
What is particularly interesting is that  $\rK^0_\rS$'s can interfere
even if they originate from a (non-identical) $\rK^0\bar\rK^0$ system
\cite{lipkin}: 
An enhancement is expected in the low-$Q$ region if one selects
the $C=+1$ eigenstate of
\beq
|\rK^0\bar\rK^0>_{C=\pm1} = \frac{1}{\sqrt 2} (|\rK^0(\vec p)\bar\rK^0(-\vec p)
\ran \pm \bar\rK^0(\vec p) \rK^0(-\vec p)\ran),
\label{12-13c}
\eeq
where $\vec p$ is the three-momentum of one of the kaons in their cms.
In the limit $\vec p\to 0\ \ (Q\to 0)$, the $C=-1\ \ (\rK^0_\rS\rK^0_\rL)$ 
state disappears and $C=+1\ \ (\rK^0_\rS\rK^0_\rS$ or $\rK^0_\rL\rK^0_\rL$)
becomes maximal.

The enhancement in $\rK^0_\rS\rK^0_\rS$ and $\rK^0_\rL\rK^0_\rL$
pairs at low $Q$ is exactly compensated by the low $Q$ suppression of the
$\rK^0_\rS\rK^0_\rL$ state, so that no BE effect is to be expected as
long as all possible final states of the $\rK^0\bar\rK^0$ system are 
considered. A full BE-like enhancement is, however, expected for the
$\rK^0_\rS\rK^0_\rS$ system by itself.

Early, low statistics results come from the hh experiment \cite{coop78}, new 
results exist from DELPHI \cite{abreu96,abreu94-2}, OPAL \cite{acto93} and 
ALEPH \cite{busk94}.  While the kaon-production radius is smaller for the hh 
experiment, it seems to agree with those measured for both charged kaons 
and pions in the e$^+$e$^-$ experiments, within the large spread of
values observed. Furthermore, the parameter $\la$ is large in agreement 
with the expectation \cite{lipkin}. 

\subsection*{4.4.3. $\La^0\La^0$ or $\bar\La^0\bar\La^0$}

An interesting generalization of the Bose-Einstein formalism used
above is to consider Fermi-Dirac interference, essentially by changing the 
sign in front of the correlator. This leads to a destructive interference
at small phase-space distance and allows to determine the emission radius
for identical fermions in a comparison of the amount of their total-spin 
$S=1$ state (destructive) to that of their $S=0$ state (constructive)
as a function of $Q$ \cite{alex95}. The method does not need a further
reference sample. It 
was applied to e$^+$e$^-$ data at LEPI in \cite{alex96-2,aleph97,delphi98}
and gives a radius of about 0.15 fm. It was, however, verified, that the
conventional method with JETSET as a reference sample gives similarly low
a radius.

\subsection*{4.4.4. (Transverse) mass dependence of the radius parameter}

The simultaneous comparison of the emission radii for pions, kaons and 
$\La$'s now suggests a decrease with increasing mass. Such a behavior has 
first been observed by NA44 in heavy-ion collisions \cite{NA44-2}.
The NA44 results can be translated into an 1/$\sqrt m_\rT$ scaling of the 
radius, in agreement with the expectations from a hydrodynamical model
\cite{csor96-2} with three-dimensional collective expansion and cylindrical 
symmetry.

In Fig.~10a, the radius parameter $r$ is shown as a function of the hadron mass
$m$ \cite{alex99} for e$^+$e$^-$ annihilation at the Z mass. The large error 
associated with $r_{\p\p}$ reflects the systematic uncertainty due to the 
choice of the reference sample. A general trend can be observed as a hierarchy
\beq
r_{\p\p} > r_{\rK\rK} > r_{\La\La} \ .
\eeq
Some effect is to be expected from kinematics, i.e. from the mass-dependent
integration limits when transforming from $R_2(\vec p_1,\vec p_2)$ in
six-dimensional momentum space to $R_2(Q)$ in one-dimensional momentum
separation \cite{Smith}. This effect is far too small, however.

 The authors \cite{alex99} show that a $1/\sqrt m$ behavior can be 
expected already from the Heisenberg principle with
\beqa
 \D p \D r &=&  mvr = \hbar c \nonumber \\
 \D E \D t &=& p^2 \D t/m = \hbar \label{12_18}\\
\mbox{and} \ \ \ r \ \ \ \ &=& \frac{c \sqrt{\hbar\D t}}{\sqrt m}\ ,\label{anne19} 
\eeqa
where $m$, $v$ and $p$ are the hadron mass, velocity and momentum and $r$
is the distance between the two hadrons. Assuming $\D E$ to only depend
on the kinetic energy of the produced particle and $\D t=10^{-24}$ sec,
independent of $m$, grants the thin solid line in Fig.~10a. The upper
and lower dashed lines correspond to an increase or decrease of $\D t$ by
$0,5\cdot10^{-24}$ sec, respectively. (The thick solid line 
corresponds to a perturbative QCD cascade using the virial theorem and
assuming local parton hadron duality (LPHD).)

\begin{figure}
\begin{center}
\epsfig{file=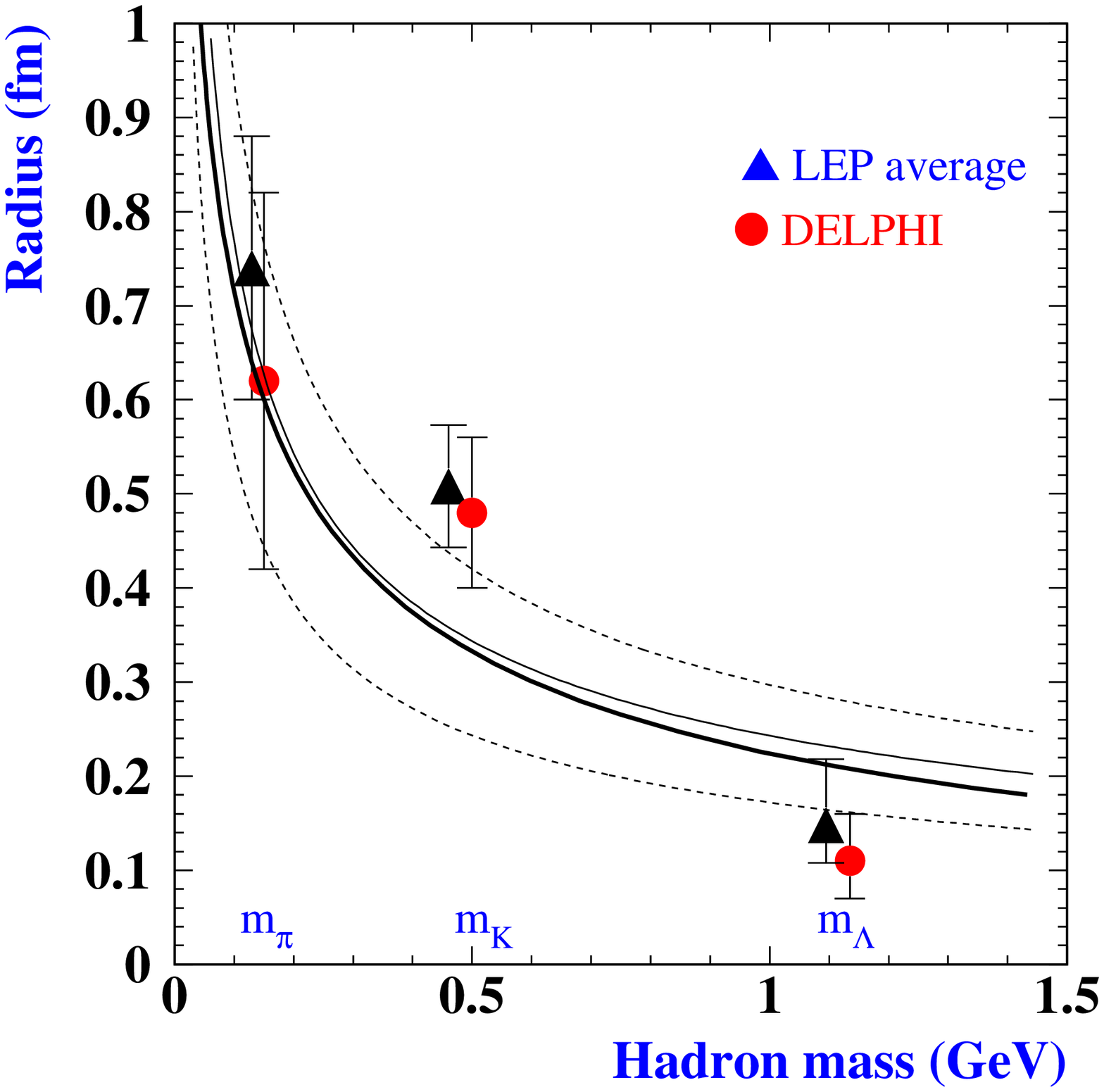,width=6cm}
\epsfig{file=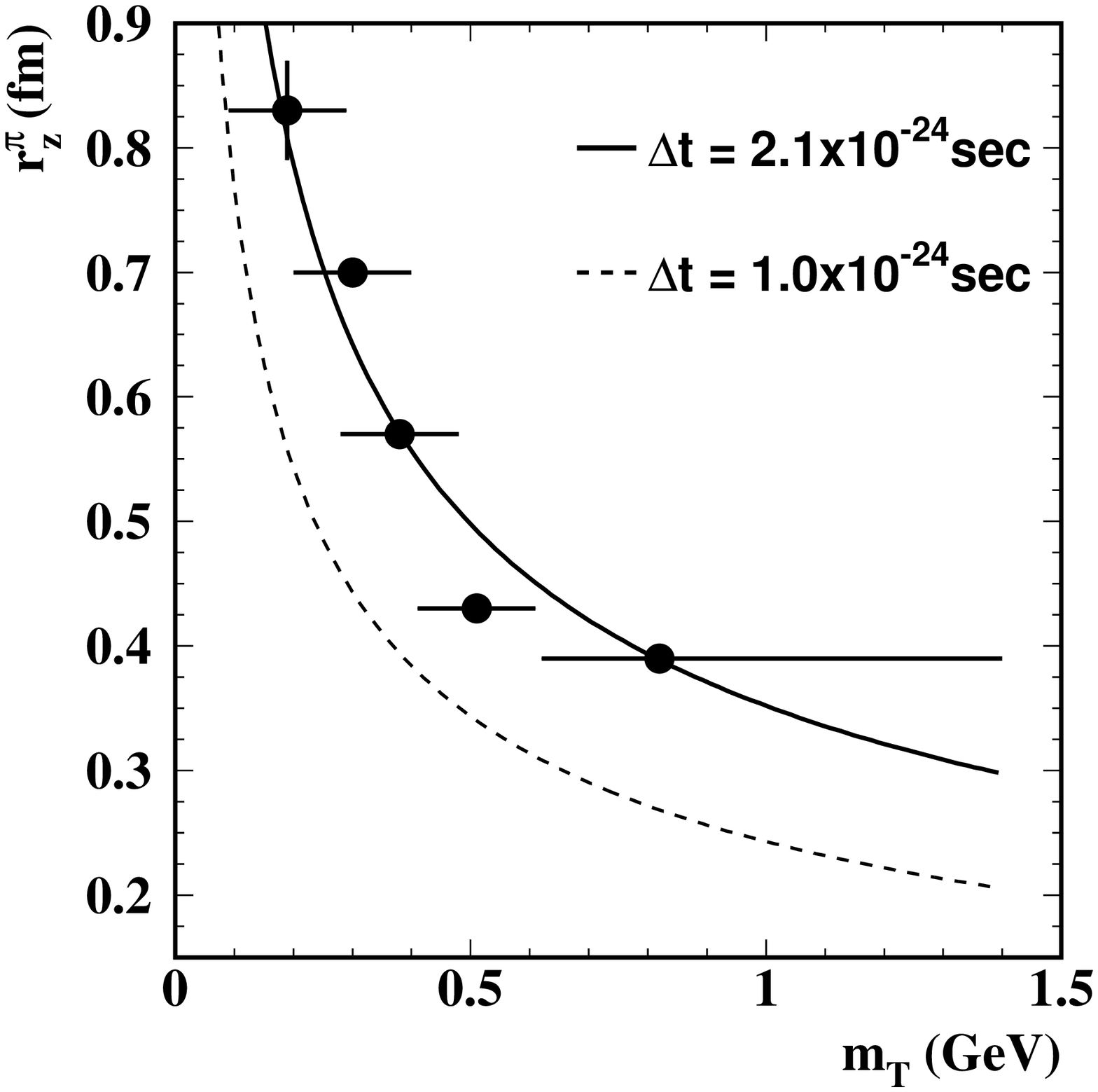,width=6cm}
\vskip-5mm
\epsfig{file=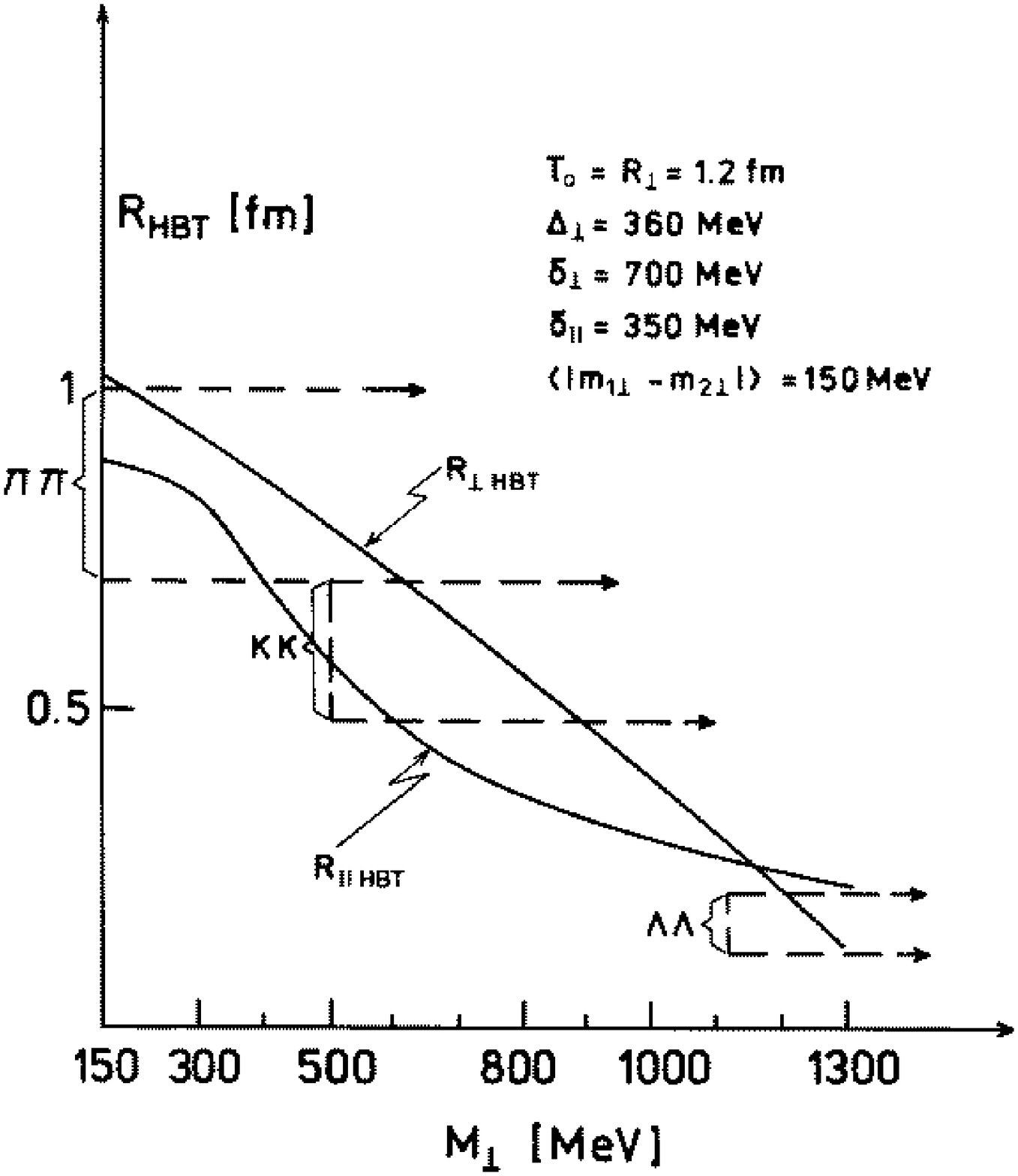,width=6cm}
\end{center}
\vs -5mm
{\small\baselineskip=12pt\ni
Figure 10. a) The radius parameter $r$ as a function of the hadron mass
$m$. b) The longitudinal emitter radius $r_z$ as a function of $m_\rT$ 
\cite{DEL}. The lines are described in the text \protect\cite{alex99}.
c) Longitudinal and transverse radius as a function of the transverse
mass $M_\bot$ of the two-particle system, compared to the $M_\bot$-threshold
values (same as data in Fig.~10a) \cite{biazal99}.\par}
\end{figure}

However, as shown in \cite{alex99}, a formula identical to (\ref{anne19})
also holds for the radius $r_\rz$ in the longitudinal direction and the
average transverse mass $\bar m_\rT=0.5$ 
$(\sqrt{m^2+p^2_{\rT 1}}+\sqrt{m^2+p^2_{\rT 2}})$. Fig.~10b shows DELPHI 
results \cite{DEL} compared to $\D t=10^{-24}$sec (dashed) and the best 
fitted value of $\D t=2.1\cdot 10^{-24}$sec (full line).

Alternatively, the transverse mass dependence can be explained by a 
generalized inside-outside cascade \cite{biazal99} assuming (i) approximate 
proportionality of four-momenta and production space-time position 
(freeze-out point) of the emitted particles $p_\m=ax_\m $
and (ii) a freeze-out time distributed along the hyperbola
$\t^2_0=t^2-z^2$ (i.e., a generalization of the so-called Bjorken-Gottfried
conditions).
From the two conditions above follows directly
\beq
a^2\t^2_0 = E^2-p^2_z=m^2_\rT
\eeq
and the generalized Bjorken-Gottfried condition
\beq
p_\m = \frac{m_\rT}{\t_0} x_\m\ .
\eeq
Using a more rigorous formulation in terms of the Wigner representation,
the authors show how this proportionality leads to an $m_\rT$ dependence
of the radius parameter.
Fig.~10c gives indeed a dependence of both the longitudinal and
the transverse radius on the transverse mass of the two-particle
system,
\beq
M^2_\bot=\bar m^2_\rT + m_{\rT 1}m_{\rT 2}\sinh^2\left(\frac{y_1-y_2}{2}\right)\ .
\eeq
For a set of ``reasonable'' model parameters \cite{biazal99}, the experimental
results (same as in Fig.~10a) are reproduced reasonably well. Note
that the experimental data are given at the threshold value of
the corresponding $M_\bot$ at which transverse momenta and rapidity 
differences are small compared to the particle masses.

The parameters are to be improved, but $\D_\bot$ is closely related to
the average transverse momentum and $\d_\bot$ has to be considerably larger
than $\D_\bot$ in the model to satisfy the uncertainty principle. Since
$\d_\bot$ corresponds to a correlation length between transverse momentum
and transverse position at freeze-out, this correlation is rather weak.
Nevertheless, it is sufficient to create a strong variation of the
transverse radius, and suggests the existence of an important ``collective
flow'', even in the system of particles produced in e$^+$e$^-$ annihilation!
Note that strong space-time momentum-space correlations are expected not
only from hydrodynamic expansion, but also from jet fragmentation.

\subsection{The multiplicity (or density) dependence} 

In nucleus-nucleus experiments \cite{111,bamb88}, the radius $r$ was 
found to increase with increasing charged-particle multiplicity $n$. 
By relating $r$ to the size of the overlap region of the two colliding 
particles, this increase can be understood in terms of the geometrical 
model \cite{117}: a large overlap should imply a large multiplicity. On 
the other hand, no evidence for a multiplicity
dependence is found in hadron-nucleus collisions at 200 GeV/c \cite{118}.
 
After some time of confusion, the $n$ dependence is now clear for
hadron-hadron collisions. At energies below $\sqrt s \approx 30$ GeV (i.e at
$\sqrt s \approx 8$ \cite{119}, 22  \cite{103} and 27 GeV \cite{agui92}) no 
$n$-dependence
is observed for $r_\rG$. At higher energies (last ref.\cite{106} and 
\cite{107}) an
$n$-dependence starts to set in and to grow with increasing
energy (see Fig.~11a). At the highest ISR energy ($\sqrt s=62$ GeV) 
the increase is about 40\% when the
density in rapidity is doubled, but at $\sqrt s=31$ GeV the
increase is still very weak. 
The result is extended to $\sqrt s=630$ GeV by UA1 \cite{alba89} 
and to 1800 GeV by E735 \cite{alex93} in Fig.~11b. 
At very large density, 
the increase of $r_\rG$ with increasing density is
shown to extrapolate well to the heavy-ion results of NA35 \cite{bamb88} 
in Fig.~11c. The effect is reproduced in thermodynamical and hydrodynamical 
models. The $\la$ parameter, on the other hand, decreases with increasing
$n$ (not shown).

\begin{figure}
\begin{center}
\epsfig{file=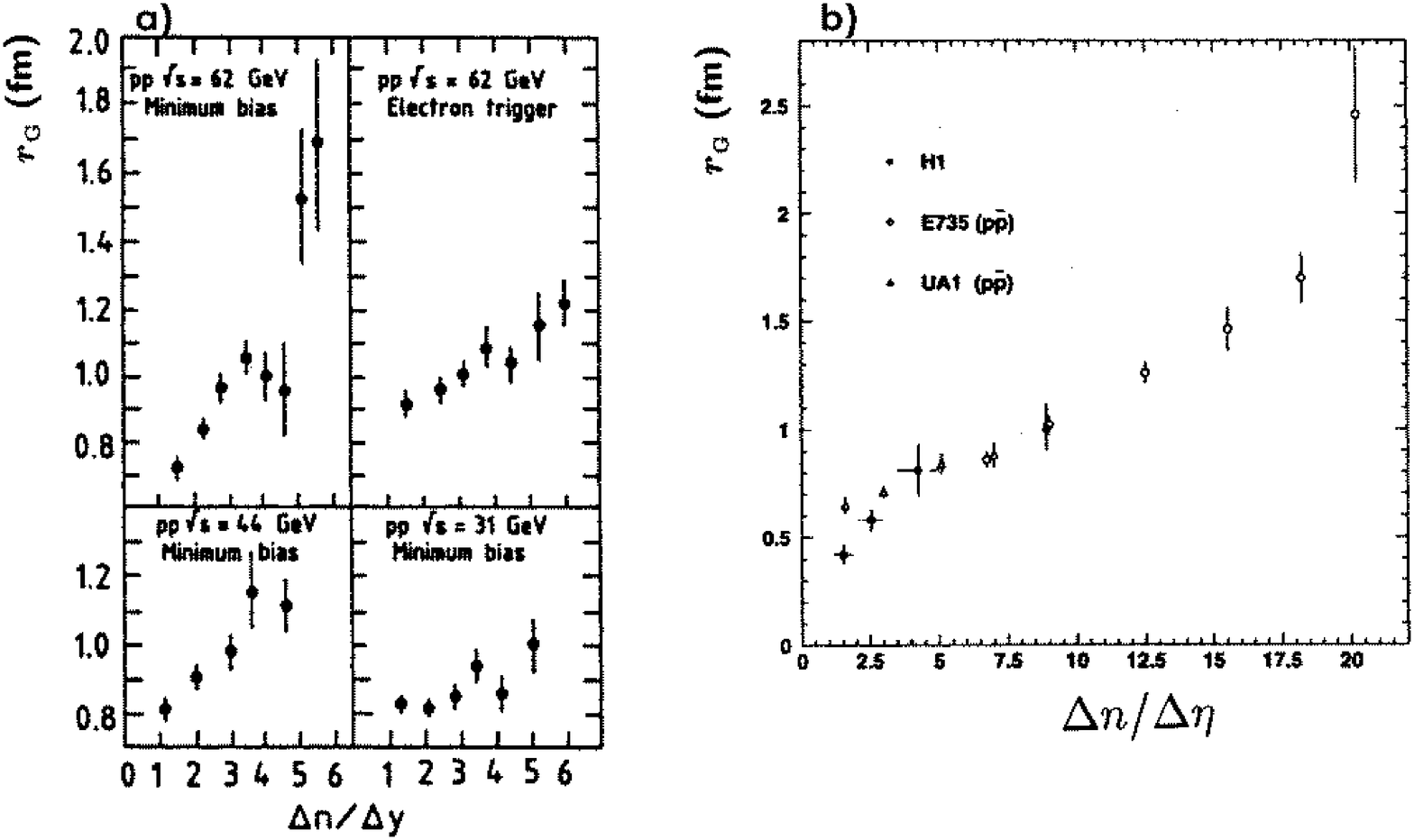,width=12cm}
\vs -2mm
\epsfig{file=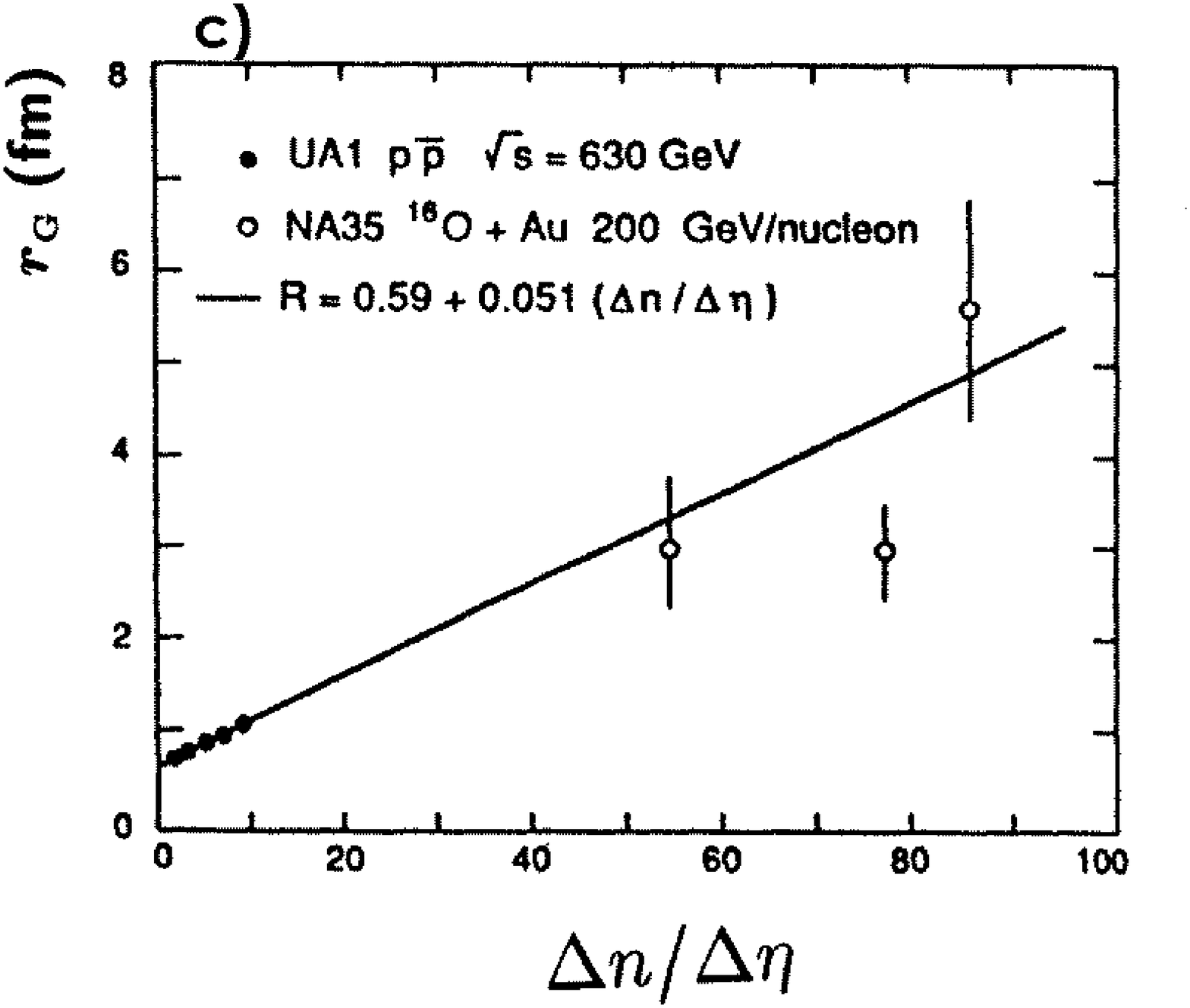,width=8cm}
\end{center}
\vs -1mm
{\small\baselineskip=12pt\ni
Figure 11 a) Radius $r_\rG$ of the pion source as a function of 
charged-particle density for the energies indicated \cite{106}, 
b) same for $\rp\bar\rp$ collisions at 630 GeV \cite{alba89} and 1800 GeV
\cite{alex93}, as well as $\re^+$p collisions at 300 GeV \cite{adloff97},
c) Comparison of $r_\rG$ as a function of charged-particle density $\D n/\D\h$
 \cite{alba89} with the results of relativistic heavy-ion collisions 
\cite{bamb88}.\par}
\end{figure}

At the low-density side, the effect is also observed in $\re^+\rp$ 
collisions by H1 \cite{adloff97} (crosses in Fig.~11b). 
The results from e$^+$e$^-$ experiments at lower energy \cite{109,110} 
were consistent with no multiplicity dependence as expected from the 
geometrical model, but also for this type of collisions a multiplicity 
dependence was finally established at higher energy \cite{alex96}
(see Fig.~12a). At 91 GeV, the radius $r_\rG$ is found to increase 
linearly with increasing multiplicity $n$, showing a
small but statistically significant 
increase of about 10\% for $10\leq n\leq40$. As for hh-collisions,
the chaoticity parameter $\la_\rG$ decreases with increasing $n$.

\begin{figure}
\begin{center}
\begin{minipage}[b]{6cm}
\epsfig{file=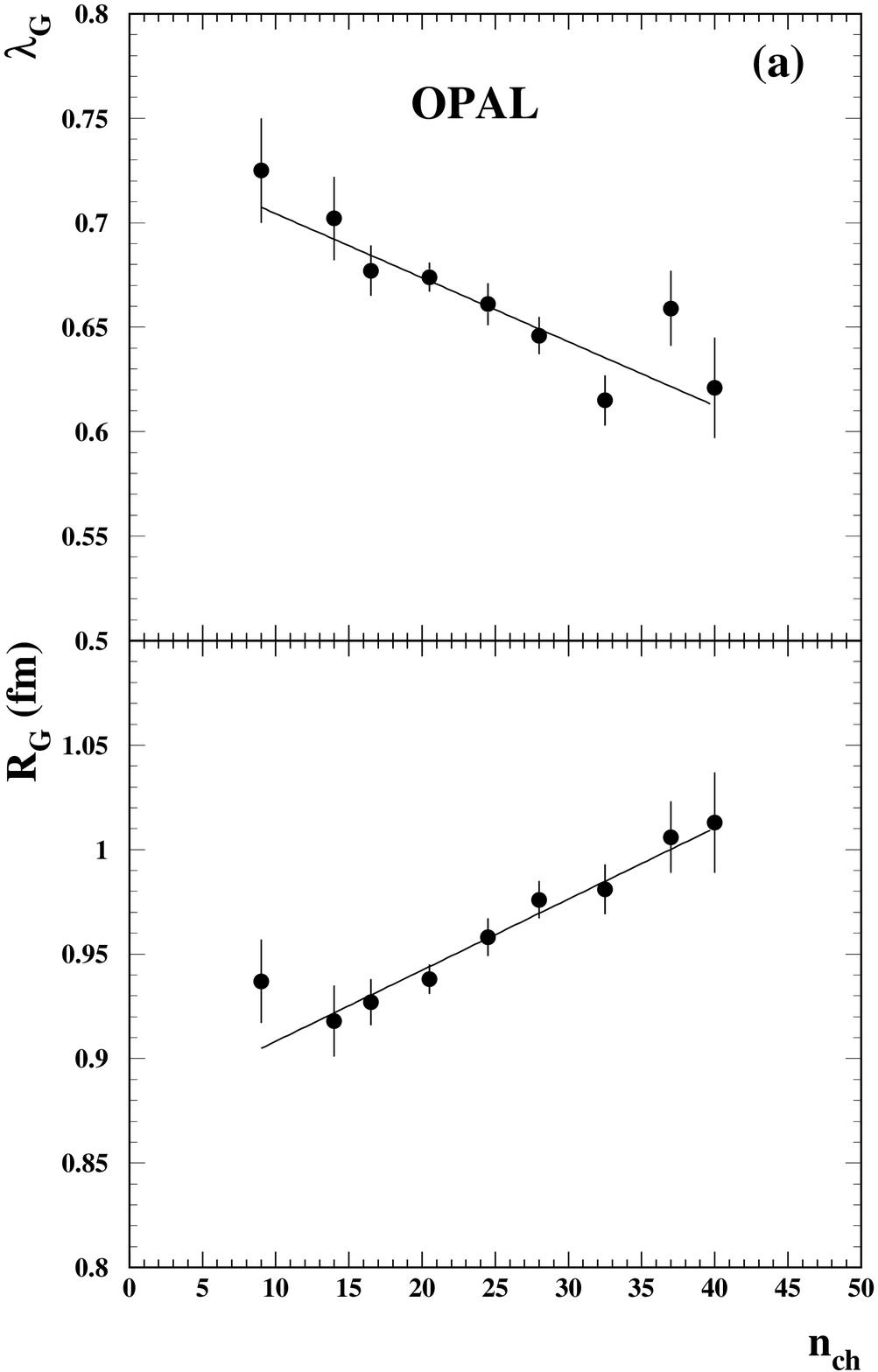,width=3.9cm}
\end{minipage}
\begin{minipage}[b]{6cm}
\epsfig{file=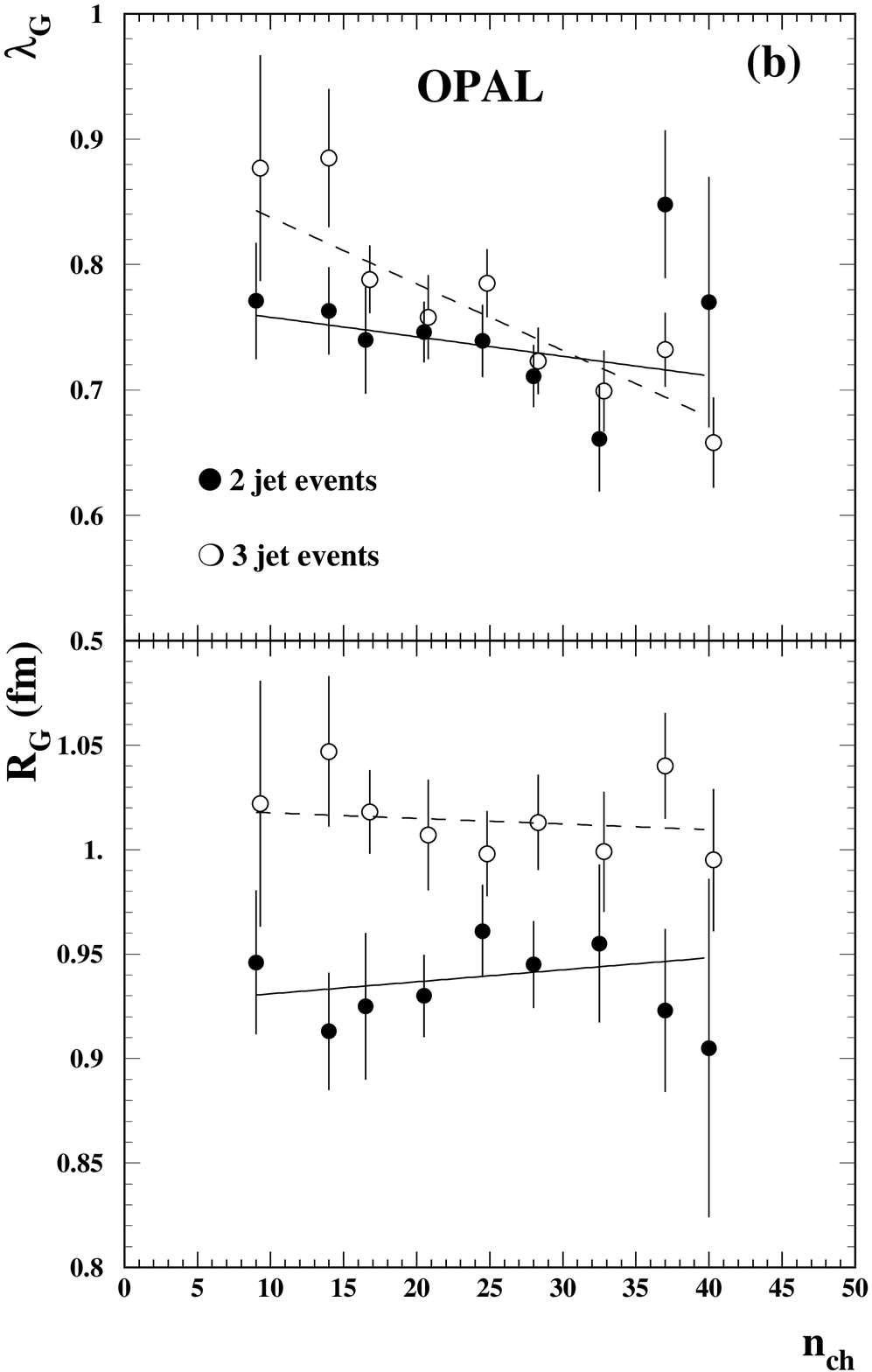,width=6cm}
\end{minipage}
\end{center}
\vs 2mm
{\small\baselineskip=12pt\ni
Figure 12.
a) Dependence of $\la_\rG$ and $r_\rG$ on the charged-particle
multiplicity $n$ for e$^+$e$^-$ collisions at the Z mass, b) same for 
two-jet events (solid points) and three-jet events (open points) \cite{alex96}.
\par}
\end{figure}

In Fig.~12b, OPAL further shows that the multiplicity dependence is strongly 
reduced in separate samples of two-jet and three-jet events, the average
value of $r_\rG$, however, being 10\% bigger for three-jet than for two-jet
events. Folding in the multiplicity difference of two- and three-jet events,
this at least partly explains the effect as due to multi-jet production
at higher energies. The decrease of $\la$ is larger in the 3-jet than in
the 2-jet sample.

As shown quantitatively in \cite{lipa96}, it is crucial to study the 
normalized cumulants $K_2(Q)$ (Eq.(11)) rather than the normalized densities 
$R_2(Q)$ (Eq.(10)) in a density dependent analysis and to correct for a 
well-defined 
multiplicity-dependent bias due to the cut in the multiplicity distribution
(the point being that $K_q\not=0$ for limited $n$, even in case of
independent emission).
 In Fig.~13a) and b) \cite{busmat}, the bias-corrected (so-called 
``internal'') cumulants are given
for UA1 as a function of the inverse rapidity density, for small and large
values of $Q$, respectively. The data show

i) a linear dependence (similar for like-charged and unlike-charged pairs),

ii) vanishing of the cumulant at large density for large $Q$,

iii) approach towards a finite limit for large density at small $Q$
(where BE correlations are expected to dominate).

\vs 2mm
\begin{center}
\epsfig{file=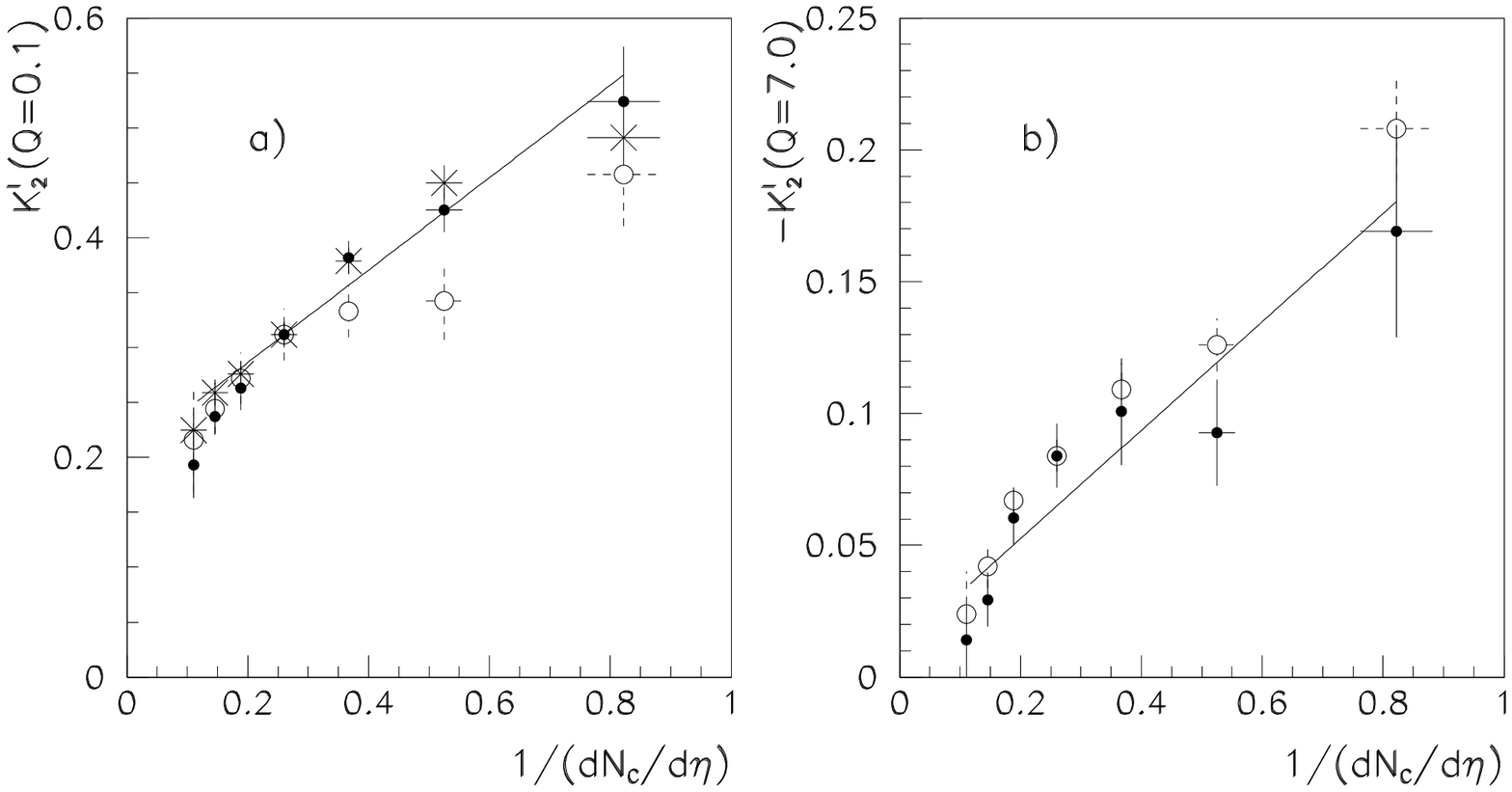,width=11cm}
\end{center}
\vs-5mm
{\small\baselineskip=12pt\ni
Figure 13. Inverse density dependence of the bias-corrected cumulant
at $Q=0.1$ GeV (Fig.~13a) and $Q=7.0$ GeV (Fig.~13b) for
like-charged pairs (full circles) and unlike-charged pairs (open circles).
The crosses in Fig.~13a correspond to $\la$-values
\protect\cite{busmat}.\par}
\vs 4mm

The large-$Q$ behavior [points i) and ii) above] is that expected from 
particle emission from $N$ fully overlapping, identical but fully
independent sources (e.g. strings). From the additivity of unnormalized 
cumulants follows immediately a dilution,
\beq
K_q^{(N)} (y_1,\dots,y_q)=\frac{N C_q (y_1,\dots,y_q)}{N^q C_1^q(y_1,\dots,
y_q)}=\frac{1}{N^{(q-1)}} K_q^{(1)}(y_1,\dots,y_q)\ ,
\eeq
where $K^{(N)}_q$ is the normalized cumulant of the $N$-source system,
while $K^{(1)}_q$ is that of an individual source (see also 
\cite{Bus,Cap,alex00}).
This results in a normalized cumulant inversely proportional to
$N$ or to the total density $\rd n/\rd y=N\rd n^{(1)}/\rd y$, 
as observed in Fig.~13b.

At small $Q$ (Fig.~13a), however, the normalized cumulant approaches
a constant different from zero at large densities. Naively, this would imply 
correlations between particles coming from different sources, which
could be interpreted as inter-source Bose-Einstein correlations, would not
a similar effect be observed for unlike-charged pairs, as well.
So also resonances play an important part.
One has to keep in mind, however, that (46) only holds for full overlap
of identical sources and that at $Q\approx 0.1$ GeV the overlap is far from
complete and the number of sources limited.

However, also in heavy-ion collisions there is evidence that $\la$
does not drop with increasing density for ever. Heavy-ion collisions
lead to $\la$ values quickly decreasing with increasing density at lower
densities (i.e. lower-$A$ collisions). In agreement with the
expectation for overlapping
independent sources, $\la$ drops  from 0.79 to 0.32 \cite{WA80}
from O-C to O-Cu, O-Ag and O-Au. In high-$A$ collisions, on the other hand,
a saturation seems to set in \cite{Biy,Paj}. For S-Pb and Pb-Pb central
collisions NA44 \cite{NA44-2} quotes $\la=0.56$ and 0.59, respectively.
Such a saturation and eventual increase of $\la$ would be expected if
the densely packed strings of a heavy-ion collision finally coalesce until 
they form a large single fireball (percolation of strings) \cite{Paj}.

\subsection{The emission function}

As has become clear from the previous sub-sections, the correlation
measurements alone do not contain the complete information on the geometrical
and dynamical parameters characterizing the evolution of the hadronic matter.
In particular, BEC are not measuring the full geometrical size of large
and expanding systems, since that expansion results in strong correlations
between space-time and momentum space. More comprehensive information can be 
provided by a combined analysis of data on two-particle correlations and 
single-particle inclusive spectra \cite{csor96,chap95-2,akke96,csor95,csor94}.

\subsection*{4.6.1. The formalism:}

In the framework of the hydrodynamical model for three-dimensionally 
expanding cylindrically-symmetric systems \cite{csor96}, the emission
function corresponds to a Boltzmann approximation of the local momentum
distribution. Within this model, the invariant single-particle spectrum of 
pions in rapidity $y$ and transverse mass $m_\rT$ is approximated by 

\begin{eqnarray}
f(y,m_\rT)&=&\frac{1}{N_{\ev}}\frac{\rd N_{\pi}}{\rd y \rd m_\rT^2} = 
\nonumber \\
  &=& C {m_\rT}^{\alpha}\cosh\eta_s\exp\left(\frac{\Delta 
\eta_*^2}{2}\right)\exp\left[-\frac{(y-y_0)^2}{2\Delta y^2}\right]
\exp\left(-\frac{m_\rT}{T_0}\right) \times \nonumber  \\ 
 && \times \exp\left\{\frac{\langle 
u_\rT\rangle^2(m_\rT^2-m_{\pi}^2)}{2T_0[T_0+(\langle u_\rT\rangle^2+\langle
\frac{\Delta T}{T}\rangle)m_\rT]}\right\}.  
\label{405-1}
\end{eqnarray}
with
\beq
\Delta y^2 = \Delta \eta^2 + \frac{T_0}{m_\rT} 
\label{405-2}
\eeq
\beq
\frac{1}{\Delta \eta_*^2} = \frac{1}{\Delta \eta^2} + 
\frac{m_\rT}{T_0}\cosh\eta_\rs, 
\label{405-3}
\eeq
\beq
\eta_\rs = \frac{y-y_0}{1+\Delta \eta^2 \frac{m_\rT}{T_0}}.  
\label{405-4}
\eeq
The width $\Delta y$ of the rapidity distribution given by (\ref{405-2}) is 
determined by the width $\Delta \eta$ of the longitudinal space-time rapidity 
$\eta$ distribution of the pion emitters and by the thermal smearing width 
$\sqrt{T_0/m_\rT}$, where $T_0$ is the freeze-out temperature (at 
the mean freeze-out time $\tau_\rf$) at the axis of the hydrodynamical 
tube, $T_0=T_\rf(r_\rT=0)$. For the case of a slowly expanding system one 
expects $\Delta \eta \ll T_0/m_\rT$, while for the case of a
relativistic longitudinal expansion the geometrical extension $\Delta 
\eta$ can be much larger than the thermal smearing (provided $m_\rT>T_0$).  

In addition to the inhomogeneity caused by the longitudinal expansion, 
(\ref{405-1}) also 
considers the inhomogeneity related to the transverse
expansion (with the mean radial component $\langle u_\rT\rangle$ of 
hydrodynamical four-velocity) and to the transverse temperature 
inhomogeneity, characterized by the quantity
\beq
\biggl\lan \frac{\Delta T}{T} \biggr\ran = \frac{T_0}{T_{\rms}}-1,
\label{405-5}
\eeq
where $T_{\rms}=T_\rf(r_\rT=r_\rT(\rms))$ is the freeze-out temperature at the 
transverse rms radius $r_\rT(\rms)$ and at time $\tau_\rf$.

The power $\alpha$ in (\ref{405-1}) is related \cite{csor96} 
to the number $d$ of dimensions in which the expanding system is 
inhomogeneous. For the 
special case of the one-dimensional inhomogeneity ($d=1$) caused by the 
longitudinal expansion, $\alpha=1-0.5d=0.5$ (provided $\Delta 
\eta^2\gg{T_0}/{m_\rT})$. The transverse inhomogeneity of the system 
leads to smaller values of $\alpha$. The minimum value of $\alpha=-1$ is 
achieved at $d=4$ for the special case of a three-dimensionally 
expanding system with temporal change of local temperature during the 
particle emission process.

The parameter $y_0$ in (\ref{405-1}) denotes the midrapidity in the 
interaction c.m.s. and can slightly differ from $0$ due to different species 
of colliding particles. The parameter $C$ is an overall normalization 
coefficient.

Note that (\ref{405-1}) yields the single-particle spectra of the
core (the central part of the interaction that supposedly
undergoes collective expansion). However, also long-lived resonances
contribute to the single-particle spectra through their
decay products. Their contribution can be determined in the 
core-halo picture \cite{csor96-2,schlei92} by the momentum dependence of the
strength parameter $\lambda(y,m_\rT)$ of the two-particle
Bose-Einstein correlation function. 
Experimentally, the parameter is however found to be approximately 
independent of $m_\rT$ \cite{beke94,agab96,agab97}. Hence this correction
can be absorbed in the overall normalization. 

The two-dimensional distribution (\ref{405-1}) can be simplified for 
one-dimen\-sional slices \cite{csor96,csor97,agab97}:

1. At fixed $m_\rT$, the rapidity distribution reduces to the approximate 
parametrization 
\beq
f(y,m_\rT) = C_m \exp\left[ -\frac{(y-y_0)^2}{2\Delta y^2}\right],
\label{405-6}
\eeq
where $C_m$ is an $m_\rT$-dependent normalization coefficient and $y_0$ 
is defined above. The width parameter $\Delta y^2$ extracted for 
different $m_\rT$-slices is predicted to depend linearly on 
$1/m_\rT$, with slope $T_0$ and intercept $\Delta \eta^2$ (cf. (\ref{405-2})).

Note, that for static fireballs or spherically expanding shells (\ref{405-6}) 
and (\ref{405-2}) are satisfied with $\Delta\eta = 0$  \cite{csor97}. Hence, 
the experimental determination of the $1/m_\rT$ dependence of the $\Delta y$ 
parameter can be utilized to distinguish between longitudinally expanding 
finite systems versus static fireballs or spherically expanding shells.

2. At fixed $y$, the $m_\rT^2$-distribution reduces to the approximate 
parametri\-zation
\beq
f(y,m_\rT) = C_y m_\rT^{\alpha} \exp\left(-\frac{m_\rT}{T_{\eff}}\right)\ ,
\label{405-7}
\eeq
where $C_y$ is a $y$-dependent normalization coefficient and $\alpha$ 
is defined as above.

The $y$-dependent "effective temperature" $T_{\eff}(y)$ can be 
approximated as
\beq
T_{\eff}(y) = \frac{T_*}{1+a(y-y_0)^2} ,
\label{405-8}
\eeq
where $T_*$ is the maximum of $T_{\eff}(y)$ achieved at $y=y_0$, and
\beq
a = \frac{T_0 T_*}{2 m_{\pi}^2(\Delta \eta^2 + \frac{T_0}{m_{\pi}})^2}
\label{405-9}
\eeq
with $T_0$ and $\Delta \eta^2$ as defined above.

The approximations (\ref{405-6}) and (\ref{405-7}) explicitly predict a 
specific narrowing of the rapidity and transverse mass spectra with 
increasing $m_\rT$ and $y$, respectively (cf. (\ref{405-2}) and 
(\ref{405-8})). The character of these variations is expected \cite{csor97} 
to be different for the various scenarios of hadron matter evolution.    

\subsection*{4.6.2. The results:}

The $\D y^2$ values obtained from fits of the NA22 data \cite{agab97} 
by (\ref{405-6}) are given as a function of $1/m_\rT$ in Fig.~14a.
A fit to the widening of the rapidity distribution (i.e. increase
of $\D y^2$) with increasing
$1/m_\rT$ by (\ref{405-2}) gives an 
intercept $\Delta \eta^2 = 1.91\pm 0.12$ and 
slope $T_0=159\pm 38$ MeV. Thus, the width of the $y$-distribution is 
dominated by the spatial (longitudinal) distribution of pion emitters 
(inherent to longitudinally expanding systems) and not by the thermal 
properties of the hadron matter, as would be expected for static or 
radially expanding sources.
Since $\Delta\eta^2$ is significantly bigger than 0,
static fireballs or spherically expanding shells, 
able to describe the two-particle correlation data in \cite{agab96}, 
fail to reproduce the single-particle spectra.

\begin{figure}[th]
\begin{center}
\begin{minipage}{6cm}
a)
\vs-15mm
\centering\epsfig{file=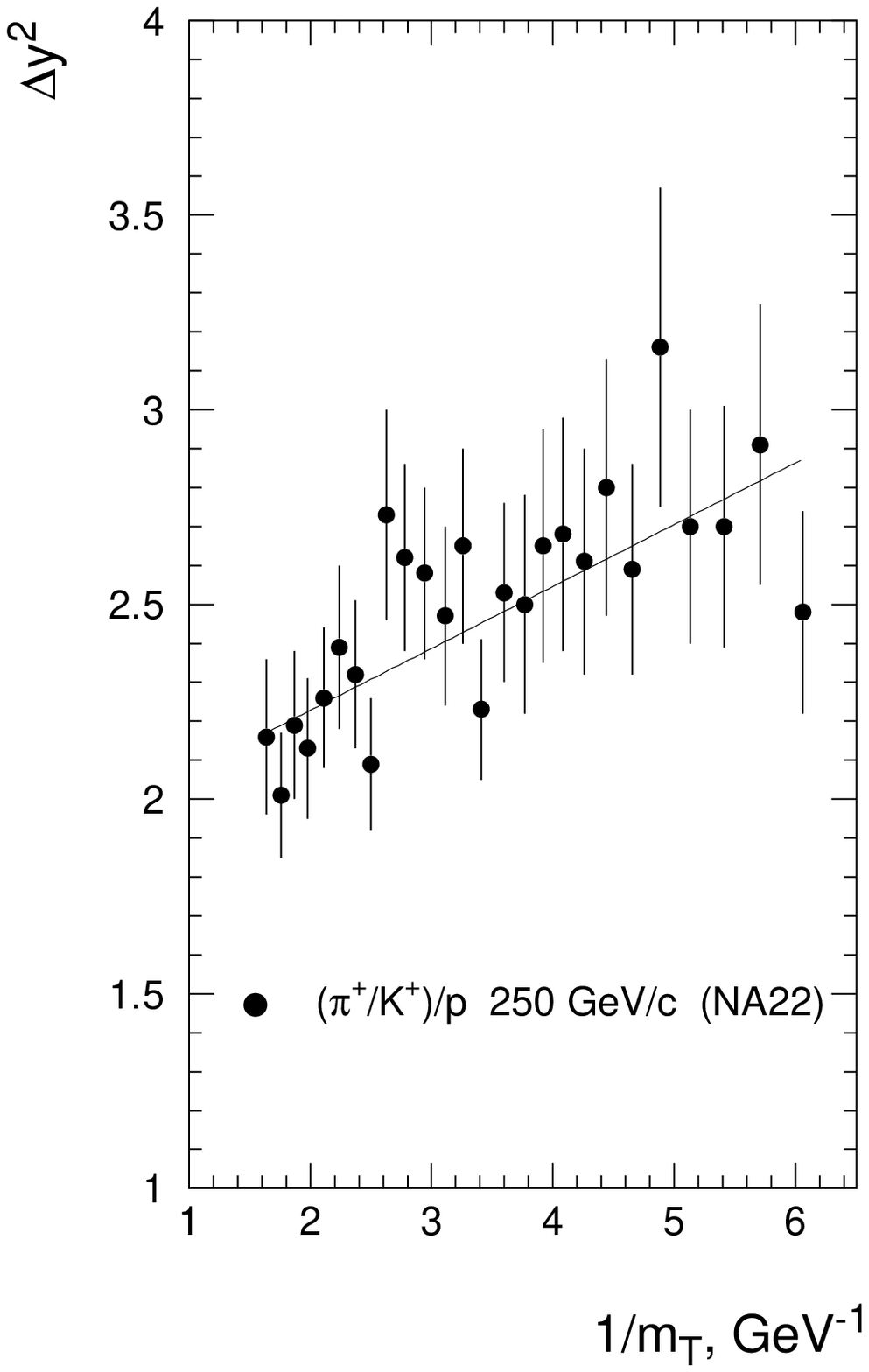,width=7cm}
\end{minipage}
\begin{minipage}{6cm}
b)
\vs-15mm
\centering\epsfig{file=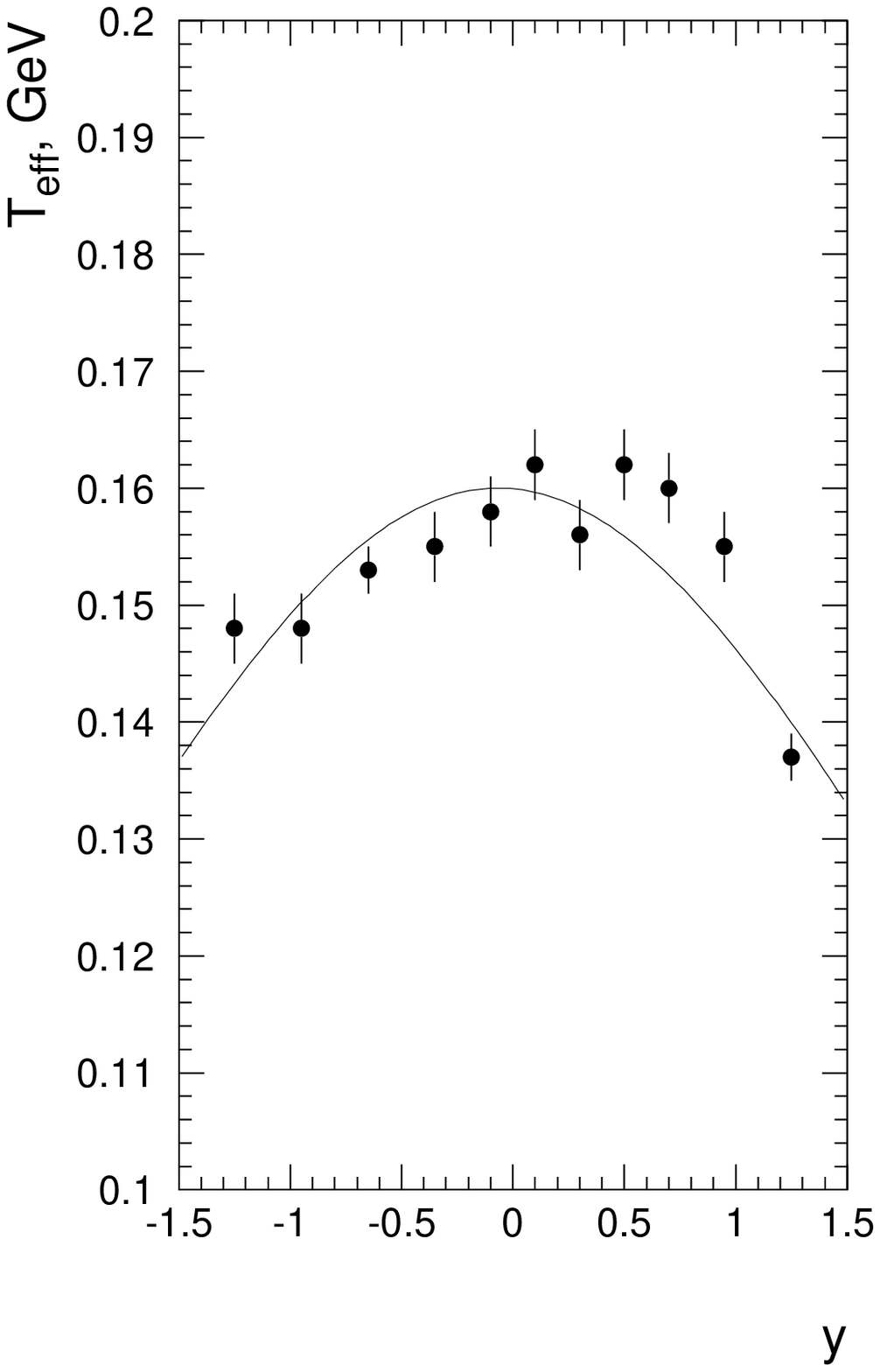,width=7cm}
\end{minipage}
\end{center}
\vs-8mm
{\small\baselineskip=12pt\ni
Figure 14.
a) The $(1/m_\rT)$-dependence of ($\Delta y)^2$ for inclusive
$\pi^-$ meson rapidity distributions at $|y|<1.5$. The straight line is 
the fit result according to parametrization (\ref{405-2}).
b) $T_{\eff}$ as a function of $y$ fitted according to parametrization 
(\ref{405-8}) \cite{agab97}.\par}
\end{figure}

The $T_{\eff}$ values obtained from fits of the same
data by (\ref{405-7}) are given as a function of $y$ in Fig.~14b.
$T_{\eff}(y)$ tends to decrease with increasing $|y|$ and approximately 
follows (\ref{405-8}) with $T_*=160\pm1$ MeV, $a=0.083\pm0.007$ and 
$y_0=-0.065\pm0.039$. Note, however, an asymmetry in the $T_\eff$ distribution
with respect to $y=0$: except for the last point, $T_\eff$ is higher in the 
meson than in the proton hemisphere.

The values of the exponential parameter $\alpha$ fitted in (\ref{405-7}) are 
near zero,
corresponding to a two-dimensional
inhomogeneity of the expanding system ($\alpha = 1-0.5d$). One concludes, 
therefore, that apart from a longitudinal inhomogeneity caused by the 
relativistic longitudinal flow, the hadron matter also has a transverse 
inhomogeneity (caused by transverse expansion or a transverse temperature 
gradient) or undergoes a temporal change of local temperature during the 
particle emission process.

\subsection*{4.6.3. The transverse direction:}

Further information on hadron-matter evolution in the transverse direction 
can be extracted from (\ref{405-1}) with parameters $\langle u_\rT\rangle $ 
and $\langle \frac{\Delta T}{T}\rangle $ characterizing the strength of 
the transverse expansion and temperature inhomogeneity.

A moderate value of the 
mean transverse four-velocity $\langle u_\rT\rangle=0.20\pm 0.07$ 
indicates that the transverse inhomogeneity is mainly 
stipulated by the rather large temperature inhomogeneity 
$\langle \frac{\Delta T}{T}\rangle =0.71\pm 0.14$. Using 
(\ref{405-5}), one infers that the freeze-out temperature decreases from 
$T_0=140\pm 3$ MeV at the central axis of the hydrodynamical tube to 
$T_{\rms}=82\pm 7$ MeV at a radial distance equal to the transverse 
rms radius of the tube.

\subsection*{4.6.4. Combination with two-particle correlations:}

Due to the non-static nature of the source, the effective size parameters
$r_\rL,r_\out,r_\side$ 
vary with the average transverse mass 
$\bar m_\rT=\frac{1}{2}(m_{\rT 1}+m_{\rT 2})$ and the average rapidity 
$Y=\frac{1}{2}(y_1+y_2)$ of the pion pair. In the LCMS the effective 
radii can be approximated \cite{csor96,csor97,csor95} by
\beq
r_\rL^2=\tau_\rf^2\Delta \eta_*^2 
\label{405-11}
\eeq
\beq
r_\out^2=r_*^2+\beta_\rT^2\Delta \tau_*^2  
\label{405-12}
\eeq
\beq
r_\side^2=r_*^2
\label{405-13}
\eeq        
with
\beq
\frac{1}{\Delta \eta_*^2}=\frac{1}{\Delta \eta^2}+\frac{\bar m_\rT}{T_0} 
\label{405-14}
\eeq
\beq
r_*^2=\frac{r_g^2}{1+\frac{\bar m_\rT}{T_0}(\langle u_\rT\rangle^2+\langle 
\frac{\Delta T}{T}\rangle)}\ \ \ , 
\label{405-15}
\eeq
where the parameters $\Delta \eta^2,T_0,\langle u_\rT\rangle$ and $\langle 
\frac{\Delta T}{T}\rangle$ are defined and estimated from the 
invariant spectra above;
$r_g$ is related to the transverse geometrical rms radius of the 
source as $r_g(\rms)=\sqrt{2} r_g$;
$\tau_\rf$ is the mean freeze-out (hadronization) time;
$\Delta \tau_\ast$ is related to the duration $\Delta \tau_f$ of pion 
emission and to the temporal inhomogeneity of the local temperature. 
If the latter has a small strength (as one can deduce from the restricted 
inhomogeneity dimension estimated above), an 
approximate relation $\Delta \tau_f \geq \Delta \tau_*$ holds.
The variable $\beta_\rT$ is the transverse velocity of the pion pair.

Using (\ref{405-11}) and (\ref{405-14}) with $T_0=140\pm 3$ MeV and 
$\Delta \eta^2 =1.85\pm 0.04$, together with $r_\rL$ fitted in different
$\bar m_\rT$ ranges, one finds a mean freeze-out 
time of $\tau_\rf =1.4\pm 0.1$ fm/c.

The transverse-plane radii $r_\out$ and $r_\side$ measured in \cite{agab96}
for the whole $\bar m_\rT$ range are: $r_\out=0.91\pm 0.08$ fm and 
$r_\side=0.54\pm 0.07$ fm. Substituting into (\ref{405-12}) and 
(\ref{405-13}), one obtains (at $\beta_\rT=0.484$c \cite{agab96}): 
$\Delta \tau_*=1.3\pm 0.3$ fm/c. Since the mean duration time of pion 
emission can be estimated as $\Delta \tau_f \geq \Delta \tau_*$, 
the data grant 
$\D\tau_f\approx \tau_\rf$. A possible interpretation is that in meson-proton 
collisions the radiation process occurs during almost all the hydrodynamical 
evolution of the hadronic matter produced.

An estimation for the parameter $r_g$ can be obtained from (\ref{405-13}) and 
(\ref{405-15}) using the quoted values of $r_\side,T_0,\langle u_\rT\rangle$
and $\langle \frac{\Delta T}{T} \rangle$. The geometrical rms transverse 
radius of the hydrodynamical tube, $r_g(\rms)=\sqrt{2}r_g=1.2\pm 0.2$ fm, 
turns out to be larger than the proton rms transverse radius.

The set of parameters of the combined analysis of single-particle spectra
and Bose-Einstein correlations in $(\p^+/\rK^+)\rp$ collisions \cite{agab97} 
is compared to that obtained \cite{ster} from averaging over Pb+Pb
experiments (NA49, NA44 and WA98) in Table~3.

\begin{table}
\begin{center}
\begin{tabular}{|l|c|c|}
\hline
Param. & NA22 & Heavy Ion \\ [-2mm]
       &      & Averaged \\
\hline
$T_0$ [MeV]      & $140\pm3$ & $139\pm6$ \\
$\lan u_\rT\ran$ & $0.20\pm0.07$ & $0.55\pm0.06$ \\
$r_g$ [fm]       & $1.2\pm0.2$ & $7.1\pm0.2$ \\
$\t_f$ [fm/c]    & $1.4\pm0.1$ & $5.9\pm0.6$ \\
$\D\t_f$ [fm/c]    & $1.3\pm0.3$ & $1.6\pm1.5$ \\
$\D\h$           & $1.36\pm0.02$ & $2.1\pm0.4$ \\
$\lan\frac{\D T}{T}\ran$  & $0.71\pm0.14$ & $0.06\pm0.05$ \\
$y_0$            & $0.082\pm0.006$ & 0 (fixed)\\
\hline
\end{tabular}
\end{center}

{\small\baselineskip=12pt\ni
Table 3. Fit parameters of the Buda-Lund hydro (BL-H) model in a combined
analysis of NA22 \cite{agab97}, NA49, NA44, WA98 \cite{ster}
spectra and correlation data.\par}
\end{table}

The temperature $T_0$ near 140 MeV comes out surprisingly similar
for hh and PbPb collisions. The geometrical radius $r_g$ and the
mean freeze-out time $\t_f$ are of course larger for PbPb than
for hh collisions, but surprising is the similarity of the duration $\D\t_f$ 
of emission in both. The fact that $\D\t_f\approx \t_f$ in hh collisions,
indicates that the radiation process occurs during all the evolution
of hadronic matter in this type of collisions. On the other hand, $\D\t_f<\t_f$
for PbPb collisions suggests that there the radiation process only sets
in at the end of the evolution. Other important differences are the
large transverse flow velocity $\lan u_\rT\ran$ and small transverse
temperature gradient in PbPb as compared to hh collisions.

\subsection*{4.6.5. The space-time distribution of $\pi$ emission:}

Figure~15a gives a reconstruction of the space-time 
distribution of pion emission points \cite{agab97}, expressed as a function 
of the cms time variable $t$ and the cms longitudinal 
coordinate $z$.
The momentum-integrated emission function  along the $z$-axis,
i.e., at $(x,y) = (0,0)$ is given by
\beq
 S(t,z) \propto \exp\left(-  {(\tau - \tau_f)^2\over 2 \Delta \tau_f^2} 
\right) \exp\left( - {(\eta - y_0)^2   \over 2 \Delta \eta^2} \right).
\label{405-19}
\eeq
\begin{figure}
\vs-1cm\hs-2.5cm
\epsfig{file=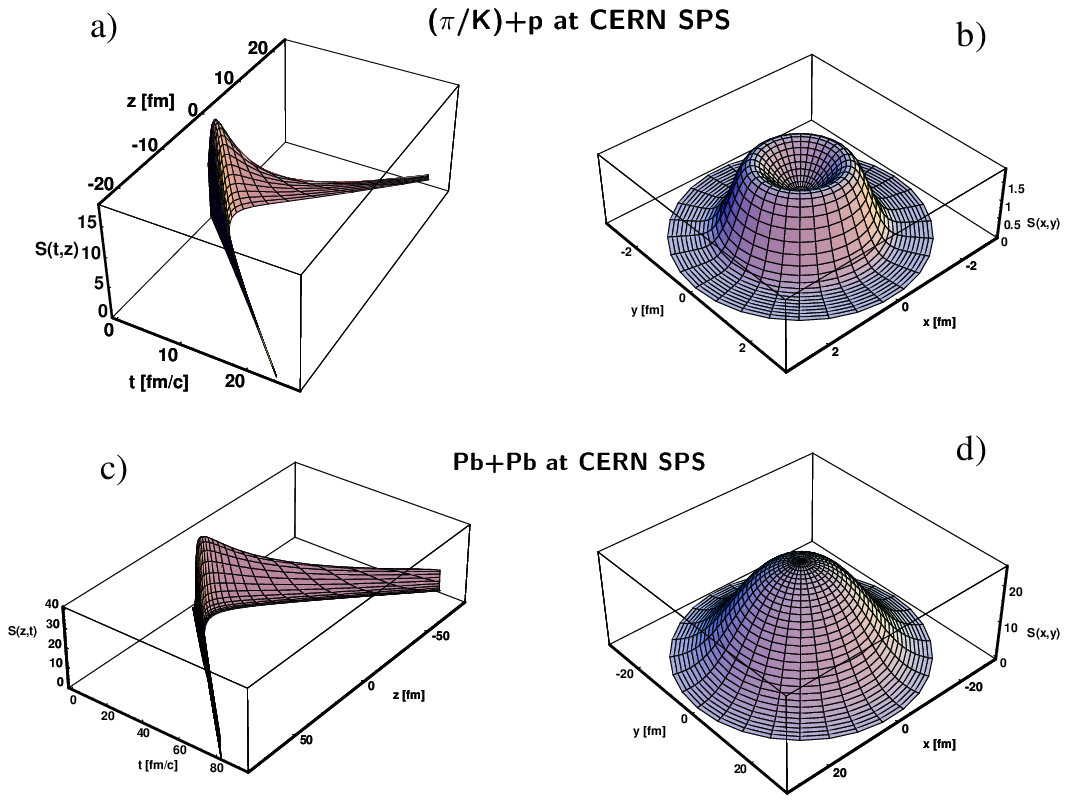,width=17.cm}

\vs -6mm
{\small\baselineskip=12pt\ni
Figure 15.
The reconstructed emission function $S(t,z)$ in arbitrary vertical units, 
as a function of time $t$ and longitudinal coordinate $z$ (left diagrams), 
as well as the reconstructed emission function $S(x,y)$ in arbitrary
vertical units, as a function of the transverse coordinates $x$ and $y$
(right pictures), for hh (upper pictures) and PbPb (lower pictures) 
collisions, respectively \protect\cite{agab97,ster,csor00}.\par}
\end{figure}

It relates the parameters fitted to the NA22 data with particle production 
in space-time. Note that the coordinates $(t,z)$, 
can be expressed with the help of the longitudinal proper-time $\tau$
and space-time rapidity $\eta$ as $(\tau \cosh\eta, \tau \sinh\eta)$.

One finds a structure resembling a boomerang, i.e., particle production takes 
place close to the regions of $z=t$ and $z=-t$, with gradually decreasing
probability for ever larger values of space-time rapidity. Although the mean 
proper-time for particle production is $\tau_\rf=1.4$ fm/c, and the dispersion
of particle production in space-time rapidity is rather small $(\Delta \eta 
= 1.36$), a characteristic long tail of particle emission is observed on 
both sides of the light-cone, giving more than 40 fm longitudinal extension 
in $z$  and 20 fm/c duration of particle production in the time variable $t$. 

An, at first sight, similar behavior is seen in Fig.~15c for 
PbPb collisions \cite{ster}. An important quantitative difference is, 
however, that particle emission starts immediately in hadron-hadron 
collision, but only after about 4-5 fm/c in PbPb collisions!

The information on $\langle u_\rT\rangle $ and 
$\langle \frac{\Delta T}{T}\rangle $ from the analysis of the transverse 
momentum distribution can be used to reconstruct the details of the 
transverse density profile. An exact, non-relativistic hydro solution was 
found \cite{csor_nucl} using an ideal gas equation of state. 
In this hydro solution, both
\beq
\langle \frac{\Delta T}{T}\rangle \gtrless \frac{m\langle 
u_\rT\rangle^2}{T_0} \ ,
\eeq
are possible. The $<$ sign corresponds to a self-similar expanding fire-ball, 
while the $>$ sign corresponds to a self-similar expanding ring of fire
(see Fig.~16). 

\vs 3mm
\cent{\epsfig{file=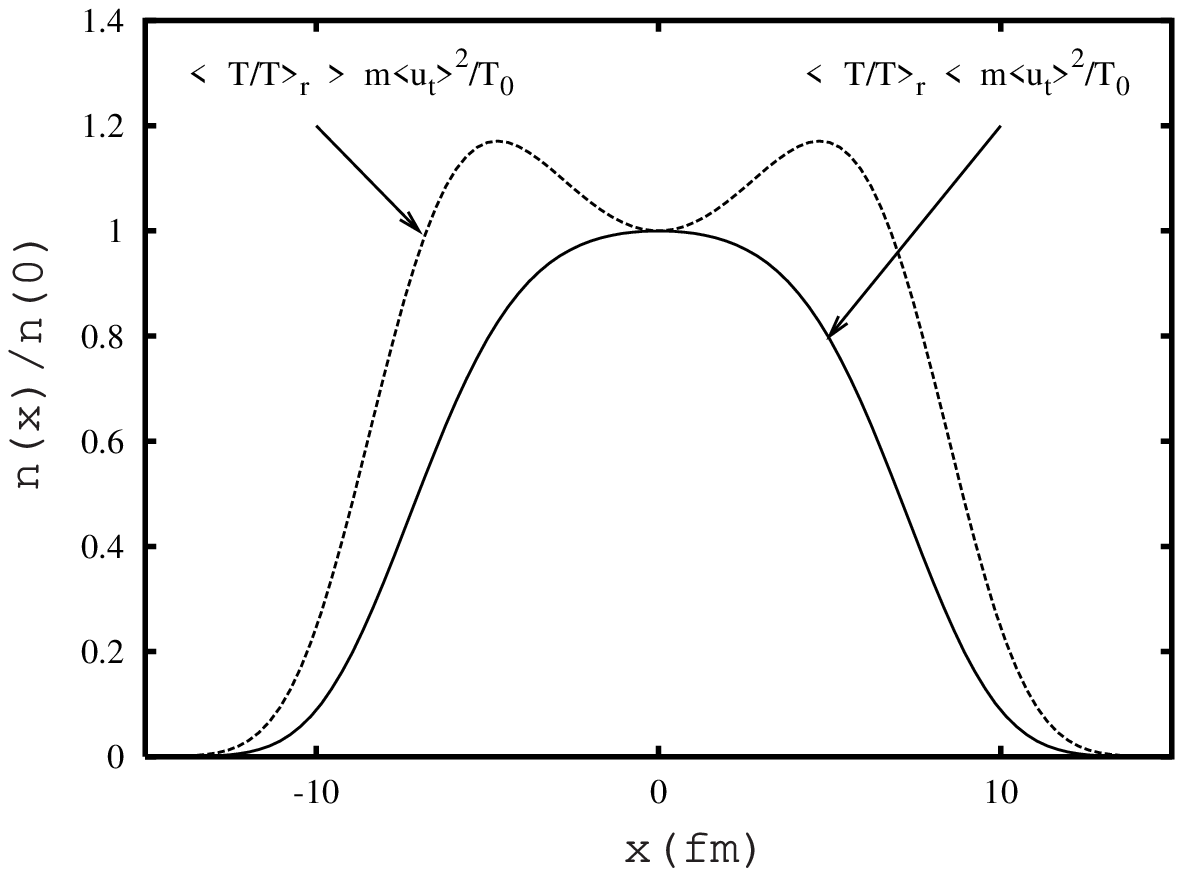,width=7.cm}}
\vs 1mm
{\small\baselineskip=12pt\ni
Figure 16. Illustration of the development of smoke-ring solutions
for large temperature gradients in exact solutions of non-relativistic
hydrodynamics \cite{csor_nucl}.\par}
\vs 4mm

Assuming the validity of this non-relativistic solution, one can reconstruct 
the detailed shape of the transverse density profile. The result looks like 
a ring of fire in the $x,y$ plane in hh interactions (Fig.~15b), while in PbPb
collisions it has a Gaussian shape (Fig.~15d). 

The formation of a ring of fire in hh collisions is due to the rather 
small transverse flow and the sudden drop
of the temperature in the transverse direction, which leads to large
pressure gradients in the center and small pressure gradients and a
density augmentation at the expanding radius of the fire-ring. 
This transverse
distribution, together with the scaling longitudinal expansion, creates an
elongated, tube-like source in three dimensions, with the density of
particle production being maximal on the surface of the tube. 

The pion emission function $S(x,y)$ for PbPb collisions, on the other hand,
corresponds to the radial
expansion, which is a well established phenomenon in heavy-ion collisions
from low-energy to high-energy reactions. This transverse
distribution, together with the scaling longitudinal expansion, creates 
a cylindrically symmetric, large and transversally
homogeneous fireball, expanding three-dimensionally with a large mean
radial component $\langle u_\rT\rangle $ of hydrodynamical four-velocity.

Because of this large difference observed for those two types of
collision, analysis of the emission function in e$^+$e$^-$ collisions is of 
crucial importance for the understanding of the actual WW overlap and has been 
started.

\subsection{The $\p^0\p^0$ system}

In a string model, unlike $\p^\pm\p^\pm$-pairs, pairs of prompt $\p^0$'s
can be emitted in adjacent string break-ups. In momentum space, the
correlation function is, therefore, expected to be wider for neutral
pions than for charged ones. Neutral pions, furthermore, do not suffer from
Coulomb repulsion. However, the detection of several $\p^0$'s in one
event requires high efficiency of $\gamma$-detection in a wide energy range
and geometrical acceptance. Furthermore, the correlation function at small 
$Q$ is strongly influenced by resonance decays as $\h\to\p^0\p^0\p^0$,
$\h'\to\p^0\p^0\h$, $K^0_\rS, f_0\to\p^0\p^0$ and other  final-state
interactions \cite{alde97}.

First evidence for Bose-Einstein correlations in $\p^0\p^0$ pairs was found 
in \cite{Esk}. In a first measurement of the radius in $\p^-$Xe interactions 
at 3.5 GeV \cite{Gris88}, the size of the $\p^0$ emission region was found 
compatible with that for charged pions.

The question was taken up again by L3 \cite{fang94}, where both $r_\rG$ and 
$\la$ are found to be on the low side when compared to the $\p^\pm\p^\pm$ 
results obtained under the same experimental conditions. The difference in 
$\la$ can at least partially be explained by the contribution of resonances. 
The difference in size parameter is $r_{\pm\pm}-r_{00}=0.150\pm0.075\pm0.068$ fm.

\subsection{Higher-Order Bose-Einstein Correlations}

\subsection*{4.8.1. The formalism}

It is convenient to use the normalized inclusive density and correlation
functions  already defined in Eqs.(10) and (11).
The normalized inclusive density for two identical pions is
\begin{equation}
R_2(1,2)=1+K_2(1,2).
\label{12-16}
\end{equation}
In the limit of a completely chaotic and static pion source, $K_2(1,2)$ 
reduces to the square of the Fourier transform $F(\vec p_1-\vec p_2,E_1-E_2)$ 
of the space-time distribution of the source, $K_2(1,2)=|F(1,2)|^2,$ where 
$\vec p_{i}$ and $E_{i}$ $(i=1,2)$ are the three-momentum and energy 
of pion $i$, respectively.

If the Gaussian parametrization is used for $|F(Q_2^2)|^2$,
then one has
\begin{equation}
K_2(Q_2^2)=|F(Q_2^2)|^2 = \exp(-r_\rG^2Q_2^2)\ \ .
\label{12-17}
\end{equation}

In terms of the $Q_{ij}$ variables and for the case of a completely 
chaotic source, the normalized inclusive three-pion density is \cite{e,l}
\begin{eqnarray}
R_3(1,2,3)&=&1+|F(Q_{12}^2)|^2+|F(Q_{23}^2)|^2
+|F(Q_{31}^2)|^2\nonumber\\
&+&2\mbox{Re}\{F(Q_{12}^2)F(Q_{23}^2)F(Q_{31}^2)\}\ ,
\label{12-18}
\end{eqnarray}
so that the genuine three-particle correlation reads
\begin{equation}
K_3(1,2,3)=2\mbox{Re}\{F(Q_{12}^2)F(Q_{23}^2)F(Q_{31}^2)\}.
\label{12-19}
\end{equation}

In general, the genuine three-particle correlation $K_3(1,2,3)$ is 
not expressed completely in terms of the two-particle correlation function 
(\ref{12-17}), but contains also new information on the phase $\f_{ij}$
of the Fourier transform of the source,
\beq
\cos\f = \frac{K_3(1,2,3)}{2\sqrt{K_2(1,2) K_2(2,3) K_2(3,1)}}\ ,
\label{12new27}
\eeq
with $\f\equiv \f_{12}+\f_{23}+\f_{31}$ being a function of $Q_{ij}$ and 
$\cos \f\to 1$ as $Q_{ij}\to 0$. Geometrical asymmetry in the production
mechanism (emission function) due to flow or resonance decays will only 
lead to small (few percent) reduction of $\cos\f$ from unity \cite{Hei}.
Equation (\ref{12new27}) is not valid for (partially) coherent sources and 
more complicated expressions are needed \cite{Hei}. If $\cos\f$ considerably
differs from unity at $Q_{ij}>0$, one can infer that partial coherence
is present (or, alternatively, that $K_3$ is suppressed due to dilution
in the case of many independent sources!).

To the extent that phase factors may be neglected and the Gaussian 
approximation would hold, $K_3$ is related to $K_2$ via the expression
\begin{equation}
K_3(Q_3^2)=2\exp(-\frac{r^2}{2}Q_3^2)=2\sqrt{K_2(Q_3^2)}
\label{12-20}
\end{equation}
with
\begin{equation}
Q_3^2\equiv Q_{123}^2=(P_1+P_2+P_3)^2-9M_{\pi}^2=
Q_{12}^2+Q_{13}^2+Q_{23}^2.
\label{12-21}
\end{equation}

\subsection*{4.8.2. Genuine three-particle correlations}

Non-zero genuine correlations up to order $q=5$ were first established by the
NA22 collaboration \cite{agab94} in terms of cumulant moments. So they must 
show up here, as well. The function $K_3(Q_3^2)+1$ is given in Fig.~17a.
A non-zero $K_3$ is indeed observed in the data for 
$Q^2_{3\p}<0.2$ (GeV/$c$)$^2$ \cite{agab95}, but not in FRITIOF. 

\vs 5mm
\begin{minipage}[t]{5.5cm}
a)
\vs -5mm\hs4mm
\epsfig{file=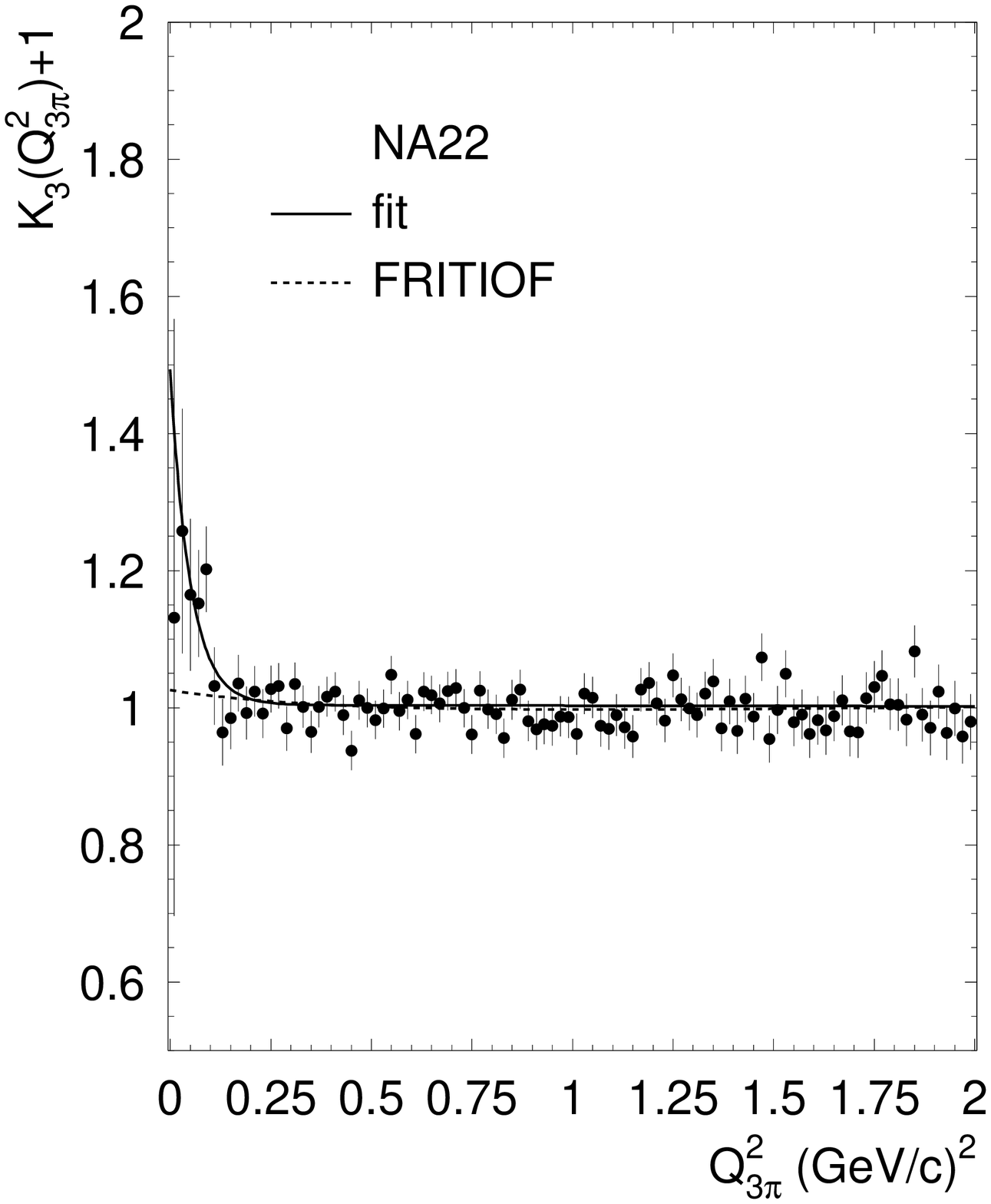,width=5.cm} 
\end{minipage}
\begin{minipage}[t]{6.5cm}
c)
\vs -5mm\hs5mm
\epsfig{file=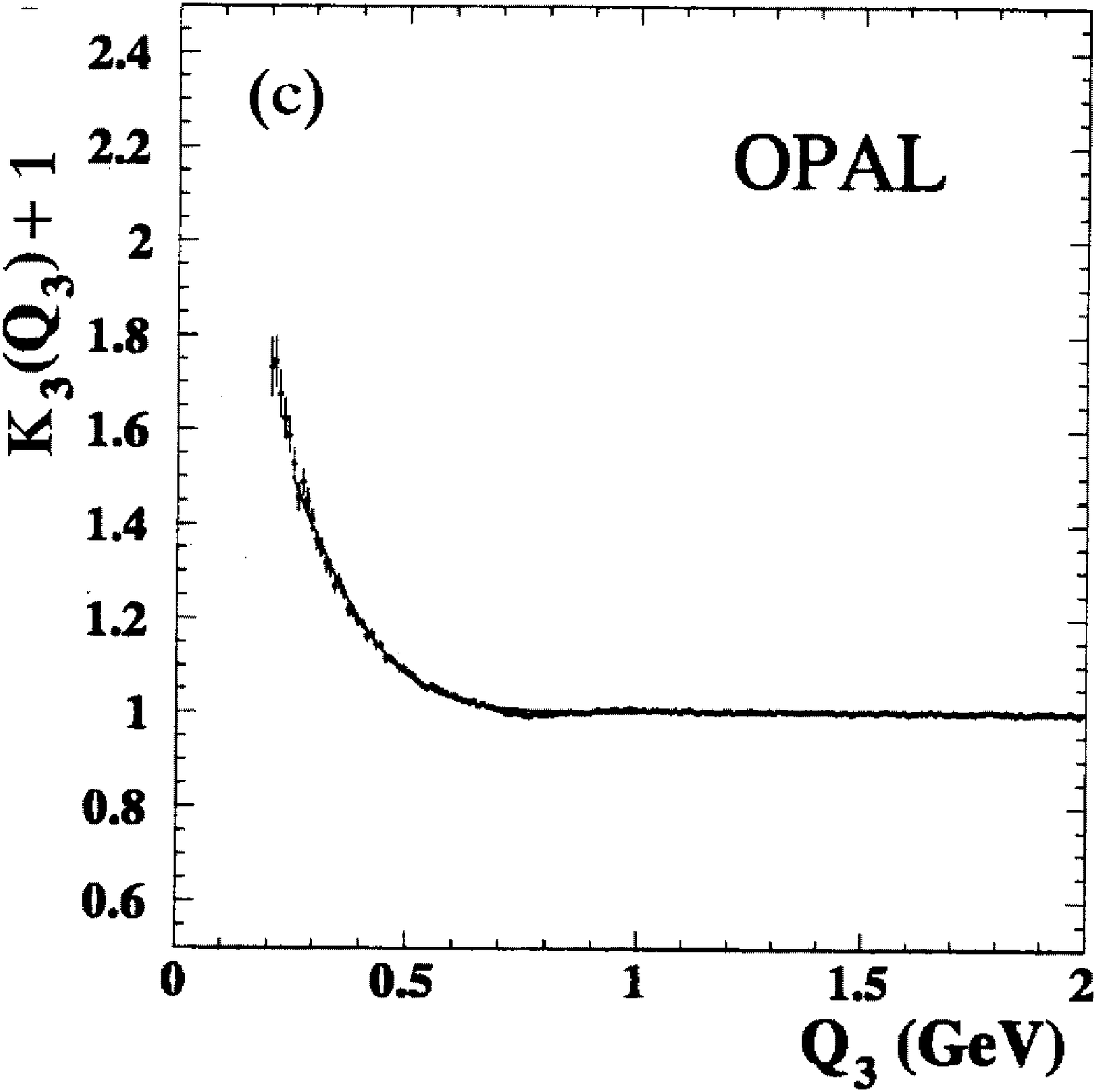,width=5.5cm} 
\end{minipage}
\vs 1mm
\hs 1.5cm b)~\vs-2cm~
\begin{center}
\epsfig{file=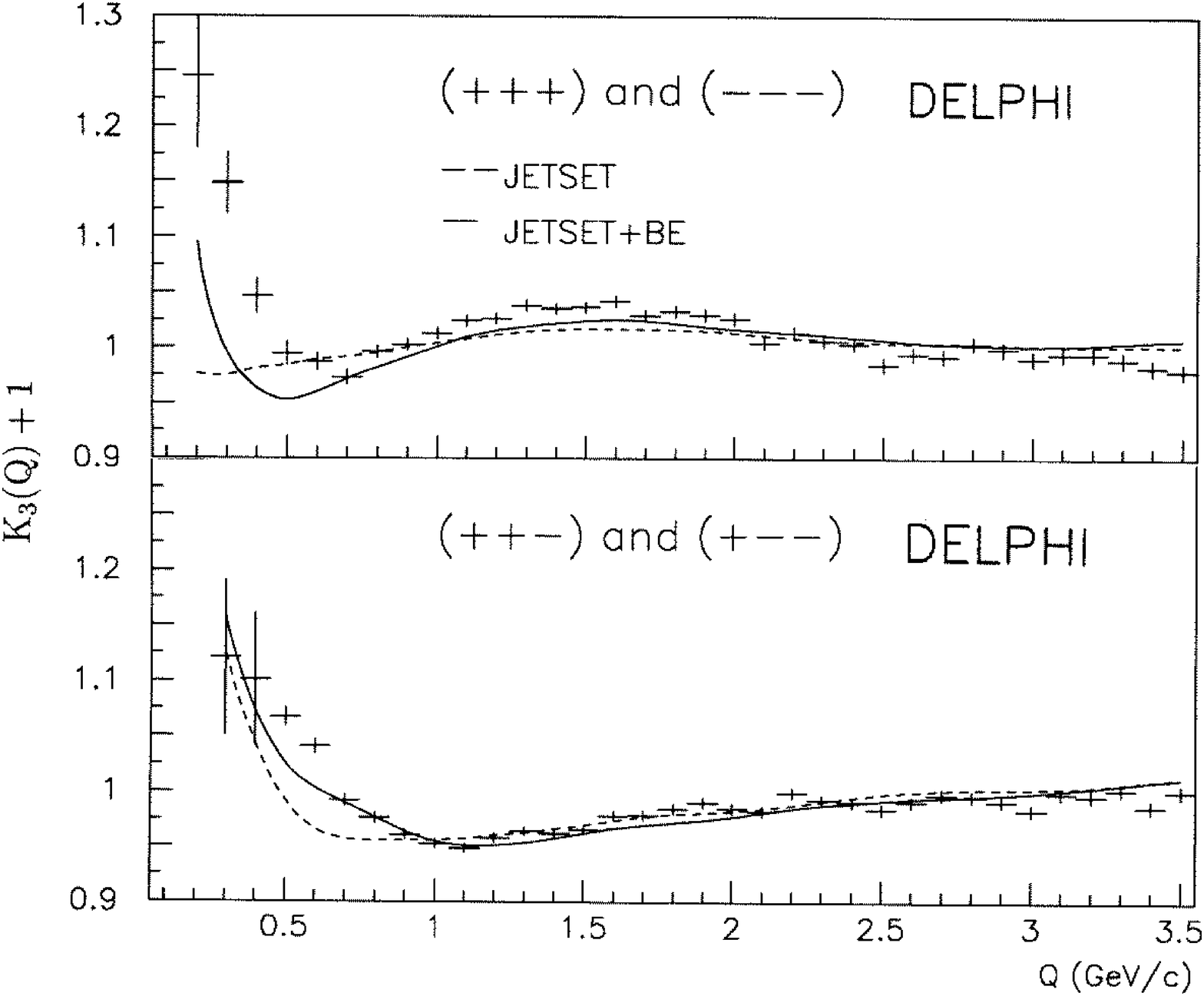,width=6.5cm}
\end{center}

{\small\baselineskip=12pt\ni
Figure~17 a) The normalized three particle correlation function
$K_3(Q_3^2)$ added to 1. The full line is the result of a fit by 
(\ref{12-31}), the dashed line corresponds to FRITIOF results \cite{agab95}.
b) The function $K_3(Q)+1$ for like-sign triplets and unlike-sign triplets. 
The predictions of JETSET without BE (dashed line) and with BE correlations 
(full line) are also shown \protect\cite{abreu95-3}; 
c) Like-sign triplets after
Coulomb correction, with a Gaussian fit (solid line) \protect\cite{acker98}.
\par}
\vs5mm

Both observations, the existence of genuine three-particle correlations
and the underestimate in JETSET are supported by DELPHI \cite{abreu95-3}. 
In Fig.~17b, the three-particle correlation function $K_3+1$ is
shown for like-charged triplets (upper) and unlike-charged triplets (lower), 
respectively, together with the prediction of JETSET with and without 
BE correlations. 
The parameters used to include the BE correlations are the same as
in the two-particle correlation study of DELPHI \cite{abreu94}. 
The model is in
reasonable agreement with the data for the $(++-)$ and $(+--)$ configurations
but underestimates the enhancement for the $(+++)$ and $(---)$ correlations.
Bose-Einstein interference in JETSET not only changes the distribution
of like-sign correlations, but also the unlike-sign ones and leads to better 
agreement with the data. 

Statistically a better evidence for genuine three-particle BE correlations
now comes from OPAL \cite{acker98}. This is shown in Fig.~17c, together 
with a Gaussian fit over the range $0.25<Q_3<2.0$ GeV, giving 
$r_3=0.580\pm 0.004\pm 0.025$ fm and $\la_3=0.504\pm 0.010 \pm 0.041$.
Within two standard deviations, the value for $r_3$ agrees with the
relation $r_3=r_2/\sqrt 2$ (see (\ref{12-20})) when compared to $r_2$ obtained 
in \cite{alex96}.

The question is, whether the observed genuine three-particle correlation
can be fully expressed in terms of the simple product of two-particle
correlation functions according to (\ref{12-20}), or whether information can
be extracted on the relative phases of (\ref{12-19}). 
If relation (\ref{12-20}) holds, the function $1+K_3(Q^2_3)$ can be
described by the parameters $r_2=0.85\pm0.01$ fm and $\la_2=0.38\pm0.02$
deduced from the fit of the normalized two-particle density $R_2(Q^2_2)$:

\begin{equation}
K_3(Q^2_3)+1=\gamma[1+2\lambda_2^{3/2}\exp(-\frac{1}{2}r_2^2Q_3^2)]
(1+\delta Q_3^2)\ .
\label{12-31}
\end{equation}
Within the errors of NA22, the resulting parameters 
$r_2$ and $\lambda_2$ do not contradict those of the two-particle 
correlations and, therefore, are consistent with Eq.(67) and, therefore,
with incoherent production of pions. DELPHI and OPAL unfortunately did not 
make use of this possibility, but L3 did \cite{L3-dal}:

Fig.~18 gives $\cos\f$ (Eq.(\ref{12new27})) as a function of $Q_3$ for
the case that the cumulants $K_2$ and $K_3$ are parametrized in terms
of a first-order Edgeworth expansion of a Gaussian. The L3 result is 
consistent with $\cos\f=1$ for all $Q_3$ and therefore with full incoherence.

\begin{figure}[h,t]
\cent{\epsfig{file=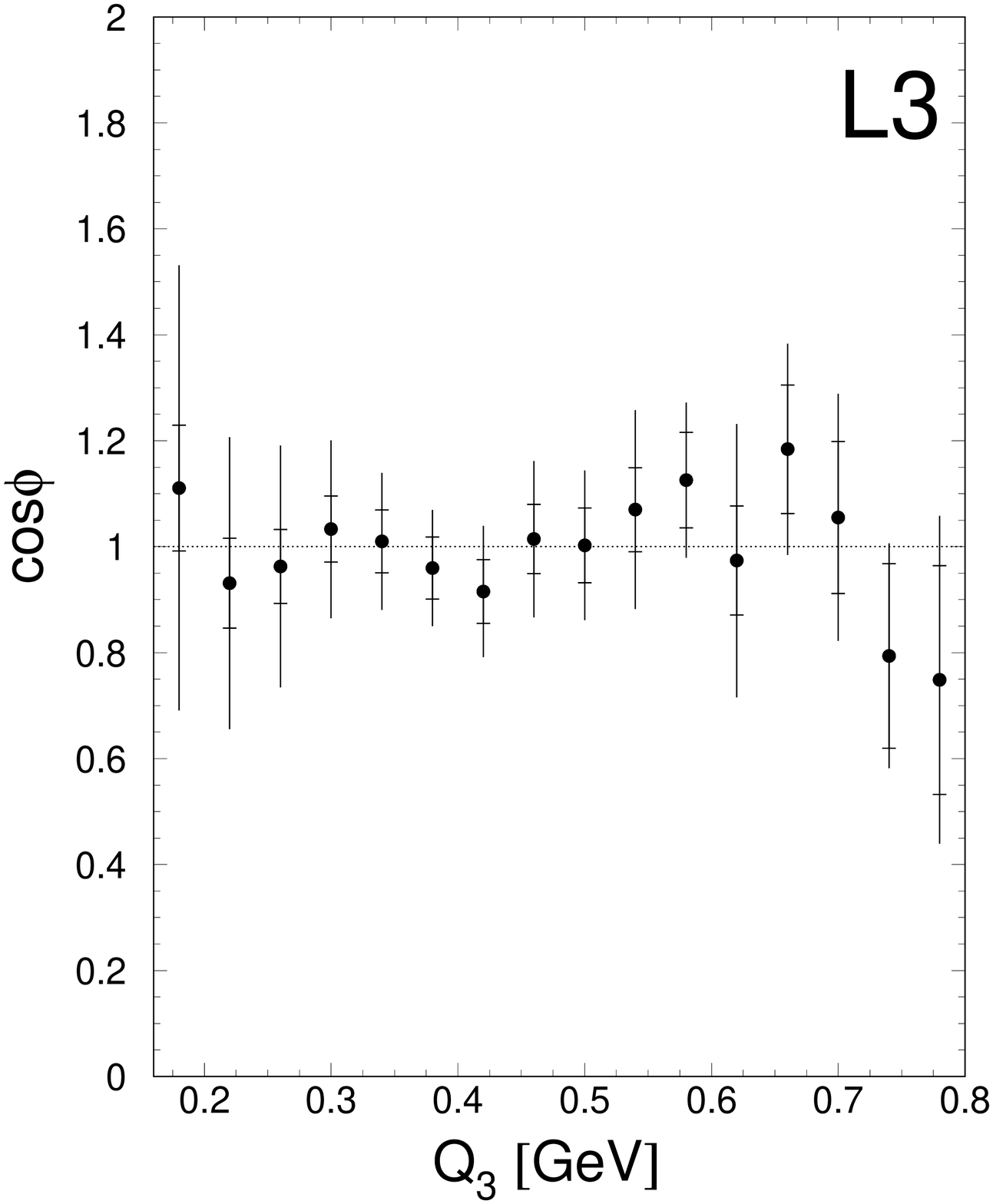,width=6cm}}

{\small\baselineskip=12pt\ni
Figure~18. $\cos\f$ as a function of $Q_3$ assuming $R_2$ is described
by the first-order Edgeworth expansion of the Gaussian \cite{L3-dal}.\par}
\end{figure}

Three-pion correlations have also been studied in heavy-ion collisions
\cite{WA98,na44} and $\lan\cos\f\ran=0.20\pm0.02\pm0.19$, i.e. no genuine 
three-particle correlations are found outside the (large) errors for SPb 
\cite{na44}. The authors interprete this result as evidence for partial 
coherence \cite{na44}.

What is particularly remarkable, however, is that the same experiment
(NA44) with the same methodology finds an average $\lan \cos\f\ran=0.85
\pm 0.02\pm0.21$ for PbPb collisions \cite{na44} and that this is supported by
a value of $\lan \cos\f\ran=0.606\pm0.005\pm0.179$ earlier reported by
WA98 \cite{WA98}. 

So,
if we trust NA44 (and I have no reason not to) and try to stick with
conventional pion interferometry, we end with a beautiful dilemma:

i) e$^+$e$^-$ collisions are consistent with fully incoherent pion production
($\cos\f\sim1$)!

ii) SPb collisions are consistent with coherent production
($\cos\f\sim 0$)!

iii) PbPb is somewhere in between!

It could not be more opposite to any reasonable expectation from conventional
interferometry \cite{thanks}.
The hint for an alternative interpretation comes from a comparison of
Eqs.~(67) and (46). What conventional interferometry calls the cosine
of a phase has in fact nothing to do with a phase. It is simply the
ratio of $K_3$ and twice $K_2^{3/2}$. It may be a challenge for the
string model to explain why this is unity for an e$^+$e$^-$ string. If that can
 be explained, the rest looks easy and very much in line with the behavior of
the strength parameter $\la$ discussed at the end of Sect.~4.5: The
ratio $\cos\f\sim K_3/2 K_2^{3/2}$ decreases with the number of independent
sources $N$ like $N^2/2N^{3/2}\propto N^{1/2}$. As $\la$ does, it decreases
with increasing atomic mass number $A$ up to SPb collisions. The saturation
or increase of $\la$ at and above this $A$ has been explained by percolation
\cite{Paj} of strings in Sect.~4.5. Exactly the same explanation can be used
to understand an increase of the ratio (not the phase!) $\cos\f$ between 
SPb and PbPb collisions.

\section{Conclusions}

In view of possible inter-W Bose-Einstein Correlations distorting
fully hadronic WW final states in e$^+$e$^-$ collisions, the state of the art 
has been summarized on Bose-Einstein correlations in Z fragmentation.
Where not (yet) available from the Z, information has been borrowed from other
types of reactions. We consider this experimental information a major
challenge to existing and future models.

1. Bose-Einstein correlations definitely exist in Z fragmentation. There is
no reason that they should not exist within a single W (intra-W BEC).
To understand possible presence or absence of BEC between pions originating
from different W's (inter-W BEC), profound knowledge of BEC in the high
statistics Z fragmentation data is obligatory.

2. The correlation is far from spherically symmetric. The correlation 
domain (defined by the lengths of homogeneity in three space directions)
is elongated along the event axis. Because of strong space-momentum
correlations, this elongation is small, however, as compared to the 
length of the total string.

3. The correlation is far from Gaussian. Good results have been obtained 
with an Edgeworth expansion, but even power-law behavior is not excluded.

4. A $1/\sqrt m_\rT$ scaling first observed in heavy-ion collisions is now
also observed in Z fragmentation and may suggest a ``transverse flow'' even 
there.

5. Dilution of correlations in high-multiplicity hh and AA collisions
suggests the lack of cross talk between neighboring strings at low
density of strings, but percolation may set in at the highest densities.

6. The full emission function in space-time can be extracted from a
combination of inclusive single-particle distributions and BE correlation
functions. So far, this has only be done in hh and AA collisions, in
the framework of a model for three-dimensional hydrodynamic expansion.
While a Gaussian shaped fire{\em ball} is observed for AA collisions,
a fire{\em tube} is observed for hh collisions. A study of e$^+$e$^-$
collisions is under way.

7. Consistently with the expectation from the string model, the radius
for $\p^0\p^0$ correlations is found to be smaller than that for
$\p^\pm\p^\pm$ correlations.

8. Genuine three-particle correlations exist in Z fragmentation and,
according to conventional interpretation, would allow to extract a ``phase'' 
unmeasurable in two-particle correlations. The resulting zero phase in
e$^+$e$^-$ is consistent with what would be expected for fully incoherent 
emission. Comparison to the results obtained for heavy-ion collisions, 
however, raises doubts on the conventional interpretation.

\section*{Acknowledgement}
I would like to thank the organizers and in particular Andrzej Bia\l as
for their kind invitation to a most pleasant meeting and 
Tamas Cs\"org\H{o} for a large number of educative discussions.


\begin{thebibliography}{99}
\bibitem{thanks}
{I would like to thank \v{S}. Todorova-Nov\'a for
encouraging discussion on this point.}
\end{thebibliography}
\end{document}